\documentclass[11pt]{article}
\usepackage{amsmath,amssymb,jheppub,graphicx}

    \usepackage{etoolbox}
    \makeatletter
    \patchcmd{\maketitle}{\@fpheader}{}
    \makeatother
    
\vspace*{0.0 mm}

\usepackage[usenames,dvipsnames]{xcolor}

\usepackage{verbatim}
\usepackage{bm} 
\usepackage{slashed}
\usepackage{mathdots}
\usepackage{color}

\usepackage{hyperref}
\def\hhref#1{\href{http://arxiv.org/abs/#1}{arXiv:#1}}

\allowdisplaybreaks[1]

\usepackage[all,cmtip]{xy}

\makeatletter
\newcommand\xleftrightarrow[2][]{
  \ext@arrow 9999{\longleftrightarrowfill@}{#1}{#2}}
\newcommand\longleftrightarrowfill@{
  \arrowfill@\leftarrow\relbar\rightarrow}
\makeatother

\newcommand{\be}{\begin{equation}}
\newcommand{\ee}{\end{equation}}
\newcommand{\bea}{\begin{eqnarray}}
\newcommand{\ea}{\end{eqnarray}}

\newcommand{\mc}{\mathcal}
\newcommand{\K}{\mathbb K}
\newcommand{\Kp}{\mathbb K^\prime}
\newcommand{\mpr}{m^\prime}
\newcommand{\nn}{\nonumber\\} 
\newcommand{\mbb}{\mathbb}
\newcommand{\sd}{  \text{sd}}

\title{Resurgence theory, ghost-instantons, and analytic continuation of path integrals}

\author[1]{G\"ok\c ce Ba\c sar}
\author[2]{Gerald~V.~Dunne}
\author[3]{and Mithat \"Unsal}
\affiliation[1]{Department of Physics and Astronomy, Stony Brook University, Stony Brook, NY 11794}
\affiliation[2]{Department of Physics, University of Connecticut, Storrs CT 06269}
\affiliation[3]{Department of Physics and Astronomy, SFSU, San Francisco, CA 94132}
\emailAdd{basar@tonic.physics.sunysb.edu}
\emailAdd{gerald.dunne@uconn.edu}
\emailAdd{unsal.mithat@gmail.com}
\abstract{
A general quantum mechanical or quantum field theoretical system in  the path integral formulation  has both real and  complex saddles (instantons and ghost-instantons).
Resurgent asymptotic analysis implies that {\it both} types of saddles   contribute to physical observables,  even if  the complex saddles are not  on the integration path i.e., the associated Stokes multipliers are zero. We  show explicitly  that instanton-anti-instanton and  ghost--anti-ghost saddles 
\textit{both} affect the expansion around the perturbative vacuum. 
 We study a self-dual model in which the analytic continuation of the partition function to negative values of coupling constant gives a 
pathological  exponential growth, but  a homotopically independent  combination of integration cycles (Lefschetz thimbles) 
 results in  a sensible theory.   These two choices of the integration cycles are tied with a  quantum phase transition.
The general set of ideas in  our construction may provide new insights into  non-perturbative QFT, string theory, quantum gravity,  and the theory of quantum phase transitions. 

}
\begin{document}
\maketitle

\section{Introduction}
\subsection{What is the physical meaning of  complex saddle points?}
\label{sec:intro}
The semiclassical weak-coupling expansion of a path integral is given by\footnote{
We normalize  the action functional as $S[\phi] = \frac{1}{g^2} \int d^d x  \; {\cal L} (\phi, \partial \phi)$ where the Lagrangian is independent of $g^2$. We denote  the action of a non-perturbative saddle as $\frac{S_k}{g^2}$ where $S_k$ is pure number.}  
\begin{eqnarray}
{\mathcal Z}(g^2)=\int {\mathcal D}\phi\, e^{- S[\phi]} \approx \sum_{{\rm saddles}\, k} F_k(g^2)\, e^{-\frac{1}{g^2}S_k} \;.
\label{pi}
\end{eqnarray}
where $F_k(g^2)$ is perturbative series around either perturbative vacuum $(k=0)$ or a non-perturbative sector.  
For systems with degenerate minima, such as the double-well and periodic potentials in quantum mechanics (QM) \cite{zinnbook}, Yang-Mills theory (see for example \cite{inst}) or semi-classically calculable deformations of Yang-Mills and ${\mathbb C}{\mathbb P}^{N-1}$ in quantum field theory (QFT)
\cite{Argyres:2012vv,Dunne:2012ae}, there are real instanton solutions, critical points of the action, giving exponentially small non-perturbative contributions $e^{-S_k/g^2}$, with $ S_k>0$.  

However, a general partition function will have critical points that are not minima but saddle points, and these will generically be complex. 
Although  the path integral  cycle is over real fields for $g^2>0$,  it can be analytically 
continued to  complex fields and couplings, providing a  global perspective on path integration \cite{pham}. 
The action, $S_k$, of a complex saddle  is also generally complex, and may even be negative. 
Such a saddle,  if taken into account na\"ively,  may lead to pathological looking  exponential growth, 
 \begin{eqnarray}
\sum_{{\rm }\, k} \sigma_k  G_k(g^2)\, e^{+\frac{1}{g^2}|S_k|} \;.
\label{pi2}
\end{eqnarray}
 of the partition function in the weak coupling limit.  But one never faces this issue for $g^2>0$, since  the integration path
 does not pass through these saddles, and they do not {\it directly} contribute to the partition function, i.e., the  Stokes multipliers $\sigma_k$ vanish,   see e.g., \cite{Witten:2010cx}.
One may na\"ively conclude that these complex saddles are irrelevant to understand the dynamics of the theory.  
In this paper, we show that it is not so, and  
explain the roles of these complex saddles (that we refer to as {\it ghost-instantons}) in the evaluation of {\it physical observables}.   Our idea is a  natural generalization to path integration (which is relevant to QM, QFT, string theory), of  the analysis of Berry and Howls for ordinary single  integrals with multiple saddles \cite{Berry}, and the ideas of Pham for ordinary multi-dimensional integrals \cite{pham}.
We call the set of all saddles  the {\it generalized instanton sector}, following the works  in $d=0$ dimensional matrix models \cite{Garoufalidis:2010ya,Marino:2007te,Marino:2008vx,Pasquetti:2009jg,Klemm:2010tm,Aniceto:2011nu,  Marino:2012zq,  Marino:2008ya}.  

We find that: 
\begin{itemize}

\item[{\it i)}]   Ghost-instantons contribute to  the large order growth of  perturbation theory (on the same footing with instantons) even though the path integration  contour
does not pass through them for $g^2>0$ (i.e. $\arg(g^2)=0$).  This is a consequence of resurgence \cite{Ecalle:1981,delabaere,Costin:2009}, 
which states that the perturbative expansion $F_0(g^2)$ in (\ref{pi}) has the information of all  (real or complex) non-perturbative saddles in coded form.\footnote{   
The property of resurgence has not been proven in general for QM via path integral formulation, or for QFT systems. Our present work provides strong evidence that observables in QM  and QFTs are resurgent functions.  A recent beautiful work by Aniceto and Schiappa demonstrates the reality of the resurgent transseries  to be a consequence of median resummation (or medianization) \cite{A-S}. 
This seems to be intimately connected by the physical mechanisms of cancellation (confluence equations)  discussed in \cite{Argyres:2012vv,Dunne:2012ae} in QFT. 
}

\item[{\it ii)}]   Upon analytic continuation of the path integral by changing $\arg(g^2)$, a ghost-instanton may become a real instanton  
  and vice versa.   At Stokes lines    one needs to  account for  the   Stokes phenomenon or   ``wall crossing". 
  \item[{\it iii)}]  We provide an example in which homotopically independent integration cycles    are in correspondence with 
   distinct  quantum  phases of an underlying theory, also see  \cite{Guralnik:2007rx, Ferrante:2013hg}. In particular, analytic continuation  and Stokes phenomenon does not permit  a ``jump" between these homotopically independent  cycles. Such a jump  is a non-analytic process, and results in  a quantum phase transition.\footnote{It is by now understood that the Stokes phenomenon itself is a continuous phenomenon \cite{Berry}. The non-analytic jump that we are referring to above cannot be obtained by Stokes phenomenon.}
\end{itemize}

In the literature, it is an accepted ``fact" that in QM systems with degenerate minima, perturbation theory leads to non-alternating asymptotic series whose large order growth is governed by 
instanton/anti-instanton $[ {\cal I}  \bar {\cal I} ]$ events, such that schematically: $a_n \sim  \frac{n!}{(2 S_I)^n}$. We show that this is  in general not always true, and that    both 
$[ {\mathcal I}  \bar {\cal I} ]$  and ghost/anti-ghost pairs $[ {\mathcal G}  \bar {\mathcal G} ]$ can be the leading  cause of divergence of perturbation theory. The ghost/anti-ghost saddles lead to an alternating asymptotic series with $a_n \sim  \frac{n!}{(2 S_{\mathcal G})^n} \sim  \frac{(-1)^n n!}{(2|S_{\mathcal G}|)^n}  $, in contrast  to the effect of  $[ {\cal I}  \bar {\cal I} ]$ events.     We show that $[ {\cal I}  \bar {\cal I} ]$
and $[ {\mathcal G}  \bar {\mathcal G} ]$
are  reflected  as branch points on the positive ($\mbb{R}^+$) and negative ($\mbb{R}^-$) axes, respectively, in the complex Borel plane, and that they {\it both} affect the expansion around the perturbative vacuum. This is an explicit realization of resurgence.  In particular, both alternating and non-alternating sub-series are present and the combination may be either alternating or non-alternating depending on the actions of the  $[ {\cal I}  \bar {\cal I} ]$  vs. $[ {\mathcal G}  \bar {\mathcal G} ]$ events, or equivalently, depending on the proximity of the associated singularity to the origin.  This behavior also emulates  the effect of both ultraviolet (UV) and infrared (IR) renormalons in asymptotically free  quantum field theories (in the sense of having singularities both on $\mbb{R}^+$ and $\mbb{R}^-$, and not in the sense of their physical origin).

Earlier successful predictions of large-order growth of perturbation theory based on purely  $[ {\cal I}  \bar {\cal I} ]$  events 
are specific to the subclass of theories without ghost-instantons, for example,  as for the periodic Sine-Gordon  potential $V(x)=\sin^2(x)$.   On the flip side,  almost all earlier cases of theories that turn out to be Borel-summable, e.g., the Sinh-Gordon potential $V(x)=\sinh^2(x)$,   only possess  ghost-instantons 
(and no real instantons)\cite{zinnbook,stone}. 
One should, however, emphasize that  both cases are {\it exceptional}. In fact,  more general potentials may have both instantons and ghost instantons, as general saddles of a path integration are typically  complex. The action associated with the saddles of path integration  are either {\it (i)} real positive; {\it (ii)} real negative, or {\it (iii)} complex conjugate pairs.  In this work, we examine a system having saddles with real positive and real negative action.  
This is the simplest QM example which  has enough structure of the general case, and   gives a glimpse into the  the inner workings of resurgence. 

In Section \ref{sec:zero}, we first introduce a zero dimensional prototype, which is the dimensional reduction of our quantum mechanics example. 
This may be viewed as the zero mode truncation  of the path integration.  The corresponding partition function is an ordinary integral, which we analyze using both analytic continuation,  resurgence and numerical methods.
 We demonstrate two remarkable results:
 \begin{itemize}
 \item[{\it a)}] The 
{\it exact} 
perturbative expansion  around the perturbative vacuum is dictated by {\it both}  non-perturbative  {\it real} and {\it complex}  saddles, i.e., their actions  and the perturbative fluctuations around them. 
\item[{\it b)}]
  {\it Late terms} of the perturbative expansion around the perturbative vacuum are dictated by the 
 {\it early terms} of the perturbative expansions around the real and complex non-perturbative saddles. This is a precise consequence of resurgence.
 \end{itemize}
 Remarkably, much of the mathematical and physical structure of this zero dimensional model appears also in the $d=1$ quantum mechanical model that we study in Section \ref{sec:one}. In particular, we demonstrate that
 both the    real and complex  saddles  contribute  to the asymptotic  growth of the perturbative expansion  around the perturbative vacuum. 
This can be viewed as a functional generalization of the Darboux theorem that underlies the resurgent structure of ordinary integrals \cite{Berry}, which says that large-orders of fluctuations about a critical point are governed by low orders about the neighboring singularities (i.e., neighboring saddles). We elaborate on this point further in Section \ref{sec:Darboux}. Our results suggest  that further  exact  resurgent relations also hold in quantum mechanical and quantum field theoretical systems.

\subsection{Ordinary integrals and path integrals  with real and complex saddles}
% We illustrate the ideas discussed above via a steepest descent analysis of the zero dimensional prototype.
%The  zero dimensional ordinary integral is the dimensional reduction of the path integral of the associated quantum mechanical system. 
To  illustrate the ideas discussed above, we consider a potential that is doubly-periodic in complex plane $\mathbb C$:
\bea
V(\phi|m)=\frac{1}{g^2}\, \sd^2(g\,\phi|m)
\label{pot}
\ea
where sd is a Jacobi elliptic function \cite{nist}, and $g^2$ is the coupling constant.  
For a doubly-periodic potential, there exists 
 both real and complex saddles,   both for  ordinary integrals as well as for  path integrals. On the other hand, 
we define the partition function for $\arg(g^2)=0$ as an integral over the real  variable    $\phi \in \mathbb R$,   and in our path integration, we integrate over fields (paths) which are manifestly real, $\phi(\tau) \in \mathbb R$. Thus, this example provides a useful testing-ground for the effects of the saddles and generalized instantons that the integration contour (cycle) {\it does}  and {\it  does not} pass through.\footnote{For a general real polynomial potential in quantum mechanics, the saddles are in general in the complex plane (through which the path integration cycle does not pass), and exceptionally, on the reals. Our analysis incorporating ghost-instantons into large-order behavior of perturbation theory also applies to such systems.}

\begin{figure}[h]
\includegraphics[scale=0.43]{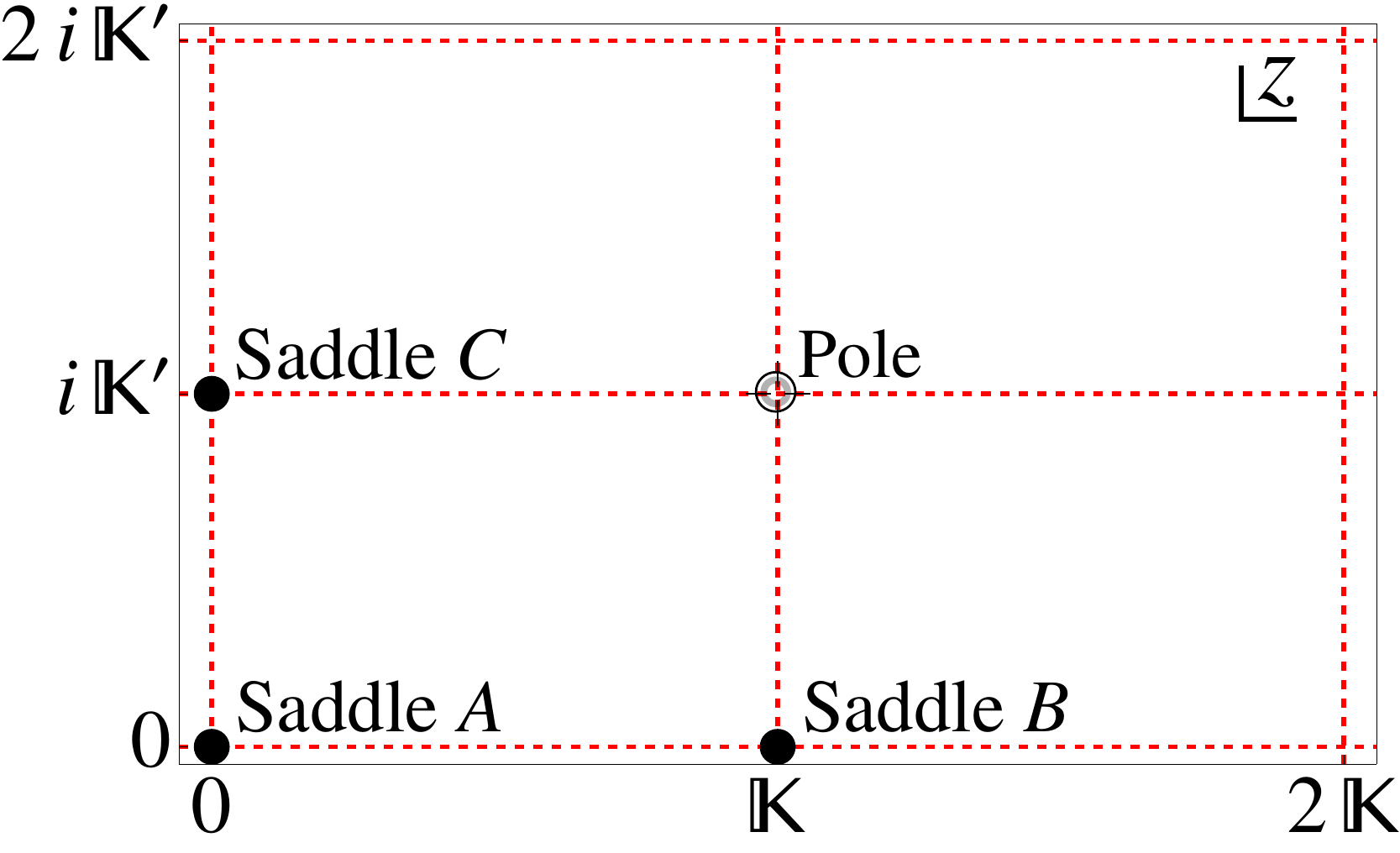} \hspace{0.5cm}
\includegraphics[scale=0.43]{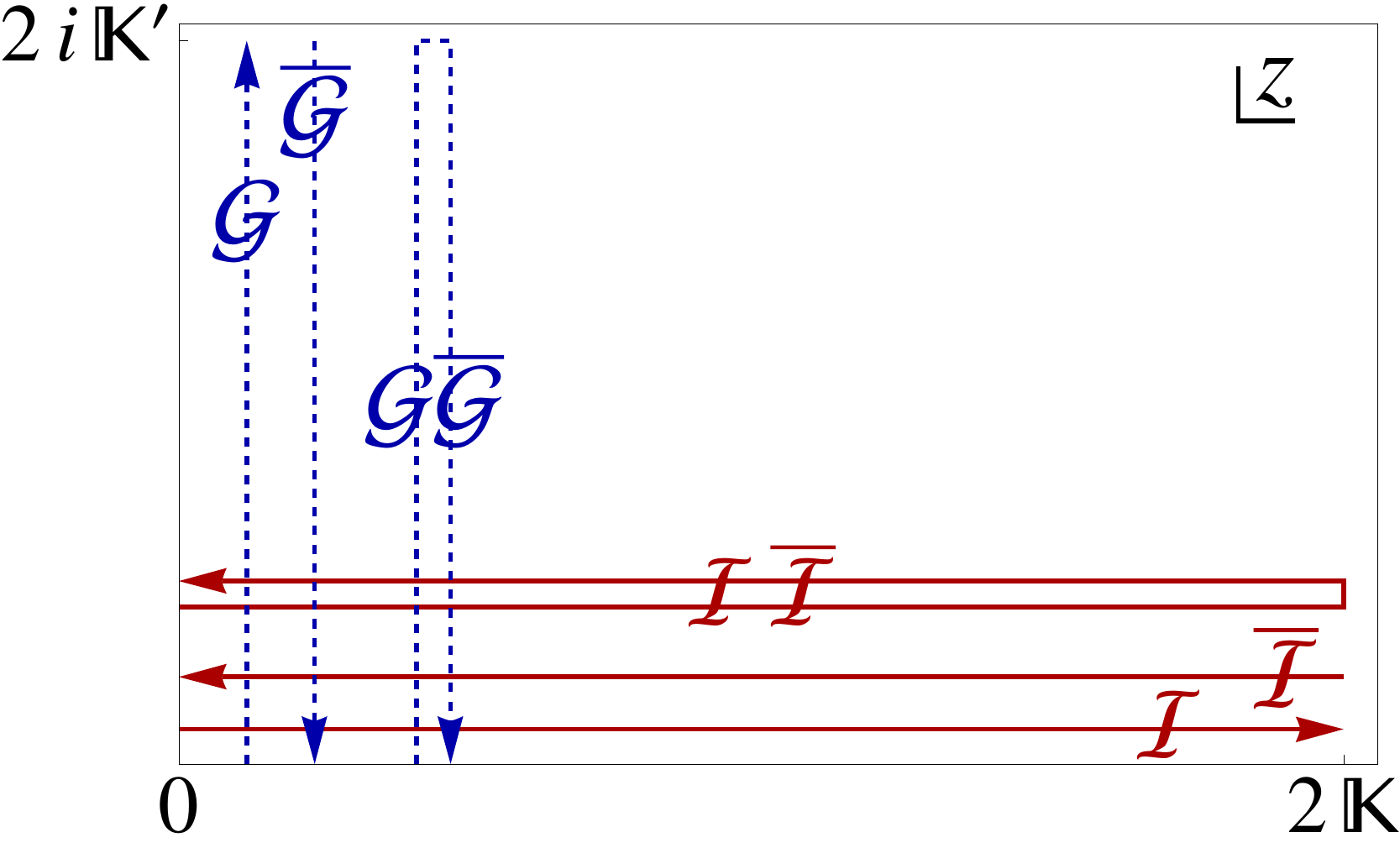}
   \caption{Doubly periodic structure of sd$^2(z|m)$ in the complex $z$ plane.  The real and imaginary periods are $2\K(m)$ and $2i\Kp(m)$ which define the fundamental torus.  
   {\bf Left:} In the $d=0$ case, studied in Section \ref{sec:zero},  the perturbative $(A)$, real $(B)$   and imaginary  $(C)$  saddles are 
 at $z_i=\{0,\K, i\Kp\}$, respectively.
 {\bf Right:}    In the $d=1$ QM example, studied in Section \ref{sec:one},  instantons and ghost-instantons are functions  which extrapolate  between $0\rightarrow  2\K(m)$ and  $0\rightarrow  2i \Kp(m)$, respectively, and $[ {\mathcal I}  \bar {\mathcal I} ]$  is a real bounce, and $[ {\mathcal G}  \bar {\mathcal G} ]$ is a complex bounce.     }
   \label{fig:rectangle}
\end{figure}

The potential (\ref{pot}) is real and periodic for $\phi\in \mathbb R$.\footnote{It is straightforward to generalize this discussion to the Weierstrass $\wp$ function. The Jacobi ellitpic function
sd$^2(z|m)$  can be expressed in terms of the Weierstrass $\wp$ function as $1/(m\mpr)[1/3(2m-1)-\wp(z+\K+i\K^\prime;\K,i\Kp)]$ where the corresponding Schr\"odinger equation is in the form of the general Lam\' e equation.}
In the complex plane, it has  two periods (one real and one imaginary) parametrized by the elliptic parameter $m\in[0,1]$ and naturally defined on a torus (see Fig \ref{fig:rectangle}). The 
real and imaginary periods are 
\bea
 2\K(m), \qquad    2i\Kp(m)=2i\K(\mpr); \qquad \qquad \mpr = 1-m,
\ea 
where $\K(m)=\int _0^{\pi/2 }\left(1-m\sin ^2(\theta )\right)^{-1/2}d\theta$  is the complete elliptic integral of the first kind \cite{nist}.
As a characteristic property of the elliptic functions it is meromorphic with a double pole at $z=\K+i\Kp$.
As a function of $m$, the potential interpolates between the trigonometric (Sine-Gordon) and hyperbolic (Sinh-Gordon) forms:
$
V(\phi|0)=\frac{1}{g^2}\, \text{sin}^2(g\,\phi)$, and $V(\phi|1)= \frac{1}{g^2}\, \text{sinh}^2(g\,\phi)
$. When $m\to 0$ the imaginary period goes to infinity, and  when $m\to 1$, the real period goes to infinity.  

Remarkably, the potential (\ref{pot}) also possesses a special ``self-duality'' property, $V(\phi|m)=-V(i\, \phi|\mpr)$. When combined with a rotation of the coupling constant $g^2\rightarrow-g^2$, this  leads to the symmetry relation:
\bea
V(\phi|m)|_{g^2}=V(\phi |\mpr)|_{-g^2}, \qquad \mpr= 1-m
\label{self_duality}
\ea
The model is ``self-dual'' in the sense that going from positive to negative $g^2$ is essentially a mapping of the model with given elliptic parameter $m$ to itself with a new elliptic parameter $\mpr=1-m$.  This important property  allows us to check certain ideas about analytic continuation, and integration cycles.  In particular, we will see that this self-duality property is not compatible with the analytical continuation of the path integral to negative $g^2$  and this is related with a quantum phase transition.

This nontrivial property of the potential also has important implications for perturbative results. Consider the perturbative expansion for \textit{any} physical quantity, such as the ground state energy,
 that can be expressed as a formal asymptotic  series for small $g^2$:
\bea
f(g^2 | m)=\sum_{n=0}^\infty a_n(m) g^{2n}
\ea
The self duality relation (\ref{self_duality}) implies that the coefficients of the perturbation series satisfy 
\bea
a_n(m)=(-1)^n a_n(\mpr)
\label{eq:selfdual}
\ea
Therefore if the series is alternating for a given $m$ it will be non-alternating for $\mpr$. When viewed na\"ively, one may conclude that  this changes its Borel summability property, but this is not the correct interpretation. In fact, there is a competition between two sub-series, one controlled by  real saddles/instantons and another controlled by   complex saddles/ghost-instantons. The question of which saddles dominate depends on the values of $m$ and ${\rm arg}(g^2)$. We demonstrate this competition explicitly below in both  $d=0$ and $d=1$. 

\section{Zero dimensional prototype}
\label{sec:zero}
\subsection{Perturbative analysis of partition function}
Our $d=0$ dimensional model can be obtained by the dimensional reduction of the $d=1$ quantum mechanics with potential (\ref{pot}). The  partition function of the zero dimensional ``field theory" is the ordinary integral: 
\bea
\mc Z(g^2 | m)=\frac{1}{g\,\sqrt{\pi}} \int_{-\K}^{\K}dz\, e^{-{1\over g^2} \,\sd^2(z | m)}\,.
\label{part}
\ea
where we interpret the exponent as the ``action functional", see (\ref{pi}). The critical point set   and  associated actions are given by 
\bea
 z_i= \{ 0,\K, i\Kp\},  \qquad S_i = \{ 0, \frac{1}{\mpr},   -\frac{1}{m}  \}, \qquad i=A, B, C
 \ea 
 Physically, these three terms correspond to the vacuum sector ($A$),  and non-perturbative  real ($B$) and complex saddles 
 ($C$).\footnote{
 In a strict sense, these are the  counterparst of the perturbative vacuum,   $[ {\mathcal I}  \bar {\cal I} ]$  and $[ {\mathcal G}  \bar {\mathcal G} ]$ saddles, respectively,  in quantum mechanics. In particular, note that $B$ and $C$ are not the counterparts of an instanton  $ {\mathcal I}$ and 
 ghost-instanton $\cal G$, but of $[ {\mathcal I}  \bar {\cal I} ]$  and $[ {\mathcal G}  \bar {\mathcal G} ]$ events. See \S\ref{sec:Darboux} for a detailed explanation.} See Figure \ref{fig:rectangle}.
 The integration contour in (\ref{part})  passes through the saddles $A$ and $B$ for $\arg(g^2)=0$, and not $C$. 
Given that sd$(0|m)=0$,  the dominant contribution along the real axis  comes from the saddle point $A$ at $z=0$.  
Below, we will establish the precise connection between perturbation theory around the perturbative vacuum (saddle $A$) and its connection to the non-perturbative real  ($B$) and complex saddles ($C$).  
We will see that these expansions are intricately entwined by resurgence.

A direct perturbative expansion of the partition function in the small $g^2$ limit around the saddle $A$ yields:
\bea
\left. \mc Z(g^2 | m)\right|_{\rm pert}&=&1-\frac{g^2}{2}(m-\frac{1}{2})+g^4 \frac{3}{32} \left(8 m^2-8 m+3\right)-g^6 \frac{15}{128} \left(16 m^3-24 m^2+18 m-5\right)\nn
&&+\dots  
\label{eq:zero-pert}
\ea
For example, the expansion coefficients for various values of elliptic parameter $m$ are:
\bea
\left. \mc Z(g^2 | 0)\right|_{\rm pert} &=&1+ \frac{g^2}{4}+ \frac{9g^4}{32}+ \frac{75g^6}{128}+ \frac{3675g^8}{2048}+ \frac{59535g^{10}}{8192} + \ldots \\\nn
\left. \mc Z\left(g^2 | 1\right)\right|_{\rm pert} &=&1- \frac{g^2}{4}+ \frac{9g^4}{32}- \frac{75g^6}{128}+ \frac{3675g^8}{2048}-\frac{59535g^{10}}{8192} + \ldots \\\nn
\left. \mc Z\left(g^2 \bigg | \frac{1}{4} \right)\right|_{\rm pert} &&  1+ \frac{g^2}{8}+ \frac{9g^4}{64}+ \frac{105g^6}{512}+ \frac{1995g^8}{4096}+ \frac{48195 g^{10}}{32768}+ \dots  \\\nn
\left. \mc Z\left(g^2 \bigg | \frac{3}{4} \right)\right|_{\rm pert} &&  1- \frac{g^2}{8}+ \frac{9g^4}{64}- \frac{105g^6}{512}+ \frac{1995g^8}{4096}- \frac{48195 g^{10}}{32768}+ \dots  \\\nn
\left. \mc Z\left(g^2 \bigg | \frac{1}{2} \right)\right|_{\rm pert} &=&1+   0 g^2 + \frac{3g^4}{32}+  0 g^6 + \frac{315 g^8}{2048}+0 g^{10} +  \ldots 
\label{d0e}
\ea
\begin{figure}
\begin{center}
\hspace{-0.5cm}
\includegraphics[width=25em]{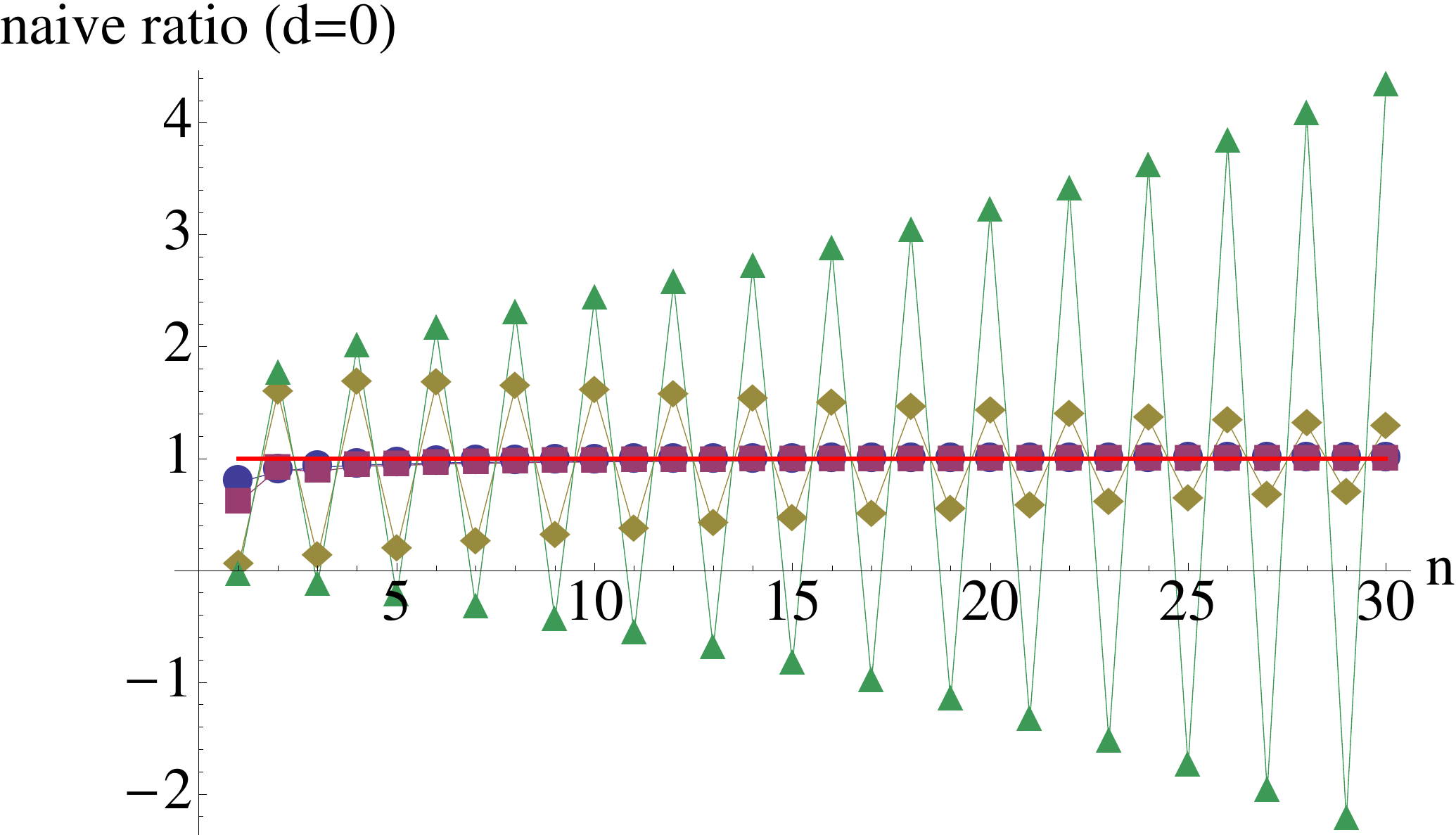}
\\
\includegraphics[width=25em]{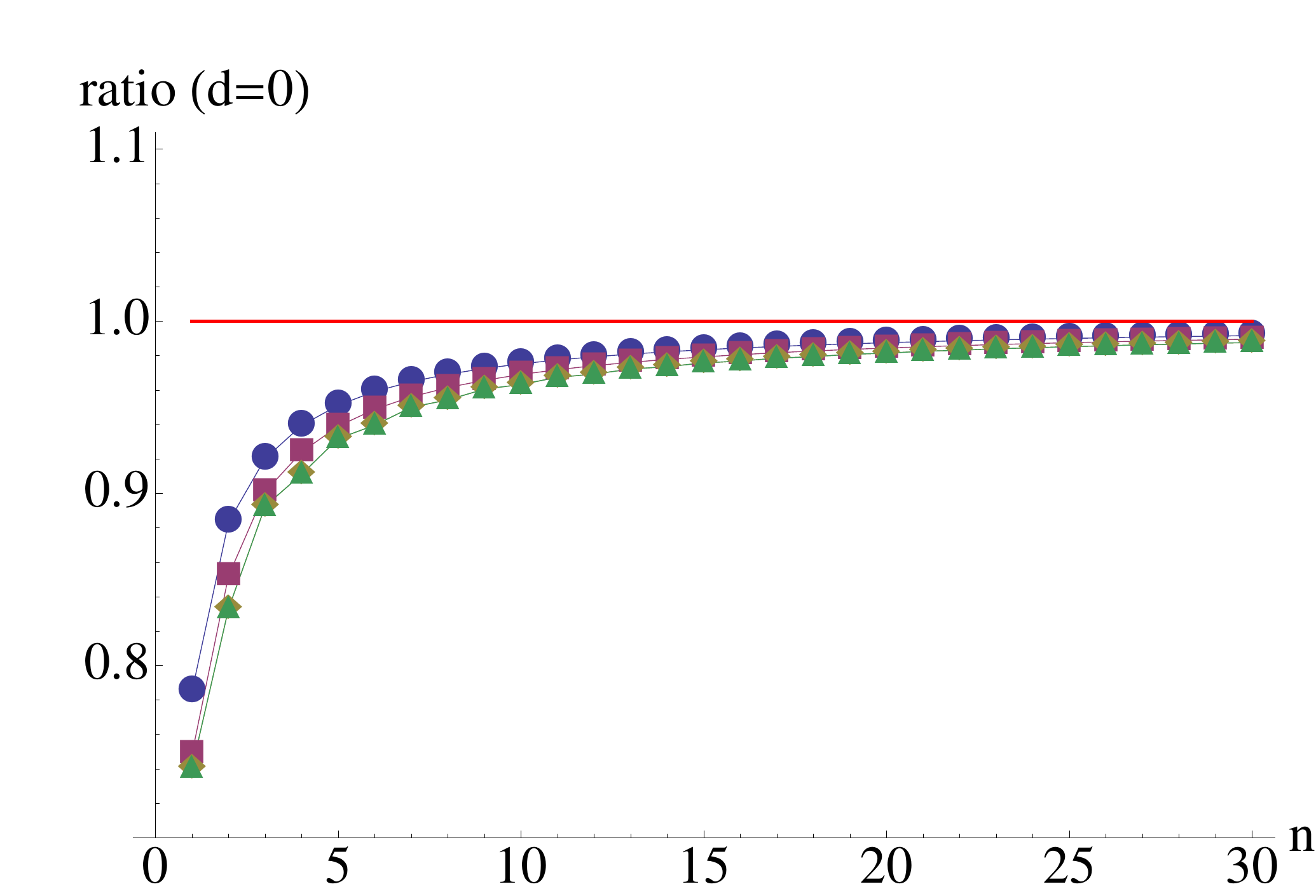}
\end{center}
\caption{{\bf Upper:}  ${a_n^{\rm actual}(m)}/{  a_n^{\rm naive}(m)} $: The ratio of the actual vacuum perturbation series coefficients to the ``naive'' prediction (\ref{eq:d0-naive}) for the large order growth which does not include the effect of complex saddle $C$.
The different curves refer to different values of the elliptic parameter $m$:  $m=0$ (blue circles), 
 $m=\frac{1}{4}$ (red squares), $m=0.49$ (gold diamonds), and $m=0.51$ (green triangles).   As $m$ approaches $1/2$ from below the agreement breaks down rapidly, showing that the contribution of the saddle $B$  by itself is not sufficient to capture the large order growth. \newline
{\bf  Lower:} The ratio ${a_n^{\rm actual}(m)}/{ a_n^{(A)}(m)} $, including the contributions to $ a_n^{(A)}(m)$ from {\it both} the  real and complex saddles, ($B$ and $C$) to leading order, as in (\ref{eq:d0-leading1}). The different curves refer to different values of the elliptic parameter $m$:  $m=0$ (blue circles), 
 $m=\frac{1}{4}$ (red squares), $m=0.49$ (gold diamonds), and $m=0.51$ (green triangles). Note the dramatically improved agreement, especially for  $m\geq 1/2$.
} 
\label{all-fig}
\end{figure}

We see that the magnitude of the coefficients  grows quickly, but we also see that when $m<\frac{1}{2}$ the coefficients are non-alternating in sign, when $m=\frac{1}{2}$ all odd terms vanish, and when $m>\frac{1}{2}$ the coefficients are alternating in sign. This reflects  the self-duality property (\ref{eq:selfdual}). 

When $m=0$ or $m=1$ the partition function is in fact a modified Bessel function in $1/g^2$, and the magnitude of the coefficients grows factorially with $n$ \cite{stone,Dunne:2012ae}. 
%\begin{figure}[htb]
%\includegraphics[scale=0.5]{d0-naive-ratio.pdf}
%\caption{The ratio of the coefficients of the perturbative expansion around saddle $A$ for $m=0$ (blue circles), $m=\frac{1}{4}$ (red squares), $m=0.49$ (gold diamonds), and $m=0.51$ (green triangles), to asymptotic form including the contribution only from the saddles $B$ to the leading order. As $m$ approaches $1/2$ from below the agreement breaks down rapidly. This figure clearly shows that the contribution of the saddle $B$ by itself is not sufficient to capture the large order growth.}
%\label{d0-naive}
%\end{figure}
For general $m$, we note that the partition function has a real saddle  point  $B$ on the integration contour at $z=\K$, at which point the ``action'' exponent is $S_B=S(m)=\frac{1}{1-m}$. A standard dispersion relation relating this instanton contribution with the large order growth of perturbation theory \cite{zinnbook} suggests the large order growth:
\bea
a_n^{\rm naive}(m)\sim \frac{(n-1)! }{\pi}(1-m)^{n+1/2}
\label{eq:d0-naive}
\ea
The ratio of the actual expansion coefficients to this naive leading form is plotted in the upper figure of Fig \ref{all-fig}, from which we see excellent agreement for small $m$, becoming worse as $m$ approaches $\frac{1}{2}$, and becoming very bad for $m>\frac{1}{2}$. Indeed, it is clear that the leading behavior in (\ref{eq:d0-naive}) cannot be correct for $m>\frac{1}{2}$, where the coefficients alternate in sign. 

The resolution of this puzzle is that there is another saddle point, $C$,  on the imaginary axis at $z= i \Kp$. This saddle also affects the perturbative expansion, even though it is  {\it not} on the integration contour. This is a consequence of Darboux's theorem, which  states that the large orders of an expansion about a critical point are governed by the behavior in the vicinity of the nearest singularity; {\it i.e.} the nearest other saddle \cite{dingle,Berry}. This is also an example of ``resurgence'' of functions represented by contour integrals, for which there is an exact intertwining relation between expansions around ``adjacent'' saddles, those that can be connected by steepest descent paths \cite{dingle,Berry} for some  $\arg(g^2)$.  

Taking into account both saddles $B$ and $C$, we obtain a corrected expression for the leading large-order growth of the perturbative expansion coefficients
\bea
a^{(A)}_n(m)&\sim  & \frac{(n-1)!}{\pi} \left[(1-m)^{n+1/2}+ (-1)^n m^{n+1/2}
\right] 
\label{eq:d0-leading1} \\
&=&  \frac{(n-1)!}{\pi} \left[\frac{1}{(S_B)^{n+1/2}}+  \frac{(-1)^n}{(|S_C|)^{n+1/2}}
\right]
\ea
The ratio of the actual expansion coefficients to this corrected leading form (\ref{eq:d0-leading1}) is plotted in the lower figure of Fig \ref{all-fig}, from which we see excellent agreement for all $m$. Notice that this corrected leading order growth also respects the self-duality property (\ref{eq:selfdual}).
This clearly confirms that the large order behavior of the perturbative expansion about saddle $A$ is governed by {\it both} the real and complex saddles, $B$ and $C$. There are two contributions, one Borel summable and the other not; the question of which one dominates is determined by the magnitude of the corresponding action, or the distance from the origin of the branch point in the Borel plane, as discussed in the next section.

\subsection{Resurgent trans-series for the partition function}
\label{sec:d0resurgence}

In this section we discuss the detailed resurgent structure of the zero dimensional partition function, and show that the improved large-order growth prediction in (\ref{eq:d0-leading1})  can in fact be improved even further by taking into account the effect of the fluctuations about the saddles $B$ and $C$. The argument is based on elementary complex analysis.

Associated with each critical point $z_i$,  there is a  unique integration cycle  ${\cal J}_i$, called a  
 Lefschetz thimble,   along which the phase remains stationary.  (A general integration contour is a linear combination of these thimbles.)
The cycle ${\cal J}_i (\theta|m)$ depending  on $\theta = \arg(g^2)$    is determined by the equation 
\be 
\label{spc}
{\rm Im} S(z) =  {\rm Im} S(z_i),  \qquad i=A,B, C
\ee
The ${\cal J}_i (\theta|m)$ cycles  are also     associated with the (downward) gradient flow of ${\rm Re} S$ where $z_i$ is a minimum of $S(z)$ along the contour. For the cases that we are interested in where $S(z)$ is a meromorphic function, the Cauchy-Riemann equations imply that the minimization problem of   ${\rm Re} S$ is same as $S$.  In general, the contours of integration deform smoothly as $\arg(g^2)$ is varied, and pass through only the associated saddle. Exactly at the Stokes lines,  these contours also pass through a subset of other saddles.  

We define the saddle point integral through  the saddle $k\in \{A,B,C\}$ (see Fig \ref{fig:rectangle}) as
\bea
\mc I^{(k)}(\xi | m)=\frac{1}{\sqrt{\pi}} \sqrt{\xi}\int_{\mc J_k}dz\,e^{-\xi\,{\text sd}^2(z|m)}
\label{saddle_z}
\ea  
where $\xi\equiv\frac{1}{g^2}$.
For  $\arg(g^2) =0^{\pm} $, the partition function (\ref{part}) is defined as the integral along the contour ${\cal C}_m$ which is  given by a linear combination of the  Lefschetz thimbles (see Fig \ref{contours}): 
\bea 
{\cal C}_m = [- \K, \K ]  =  [- \K + i \Kp,  - \K ]  \cup  [ - \K ,   \K  ] \cup [\K,  \K + i \Kp]  = 
 \left\{ \begin{array}{l} 
   {\cal J}_A(0^{-}|m) + {\cal J}_B (0^{-}|m)  \cr
     {\cal J}_A(0^{+}|m) - {\cal J}_B (0^{+}|m)  . 
\end{array} \right. \nn
\label{cycle-C}
\ea
This contour is along the real axis, and   the flip of the  sign of the ${\cal J}_B$  at $\theta=0$ is  Stokes phenomenon. 
The other two integration cycles,   $\mc J_B$ and $\mc J_C$ are parallel to the imaginary axis  for $\arg(g^2) =0$  in the 
vicinity of the   corresponding saddle.

\begin{figure}[htb]
\center
\includegraphics[scale=0.55]{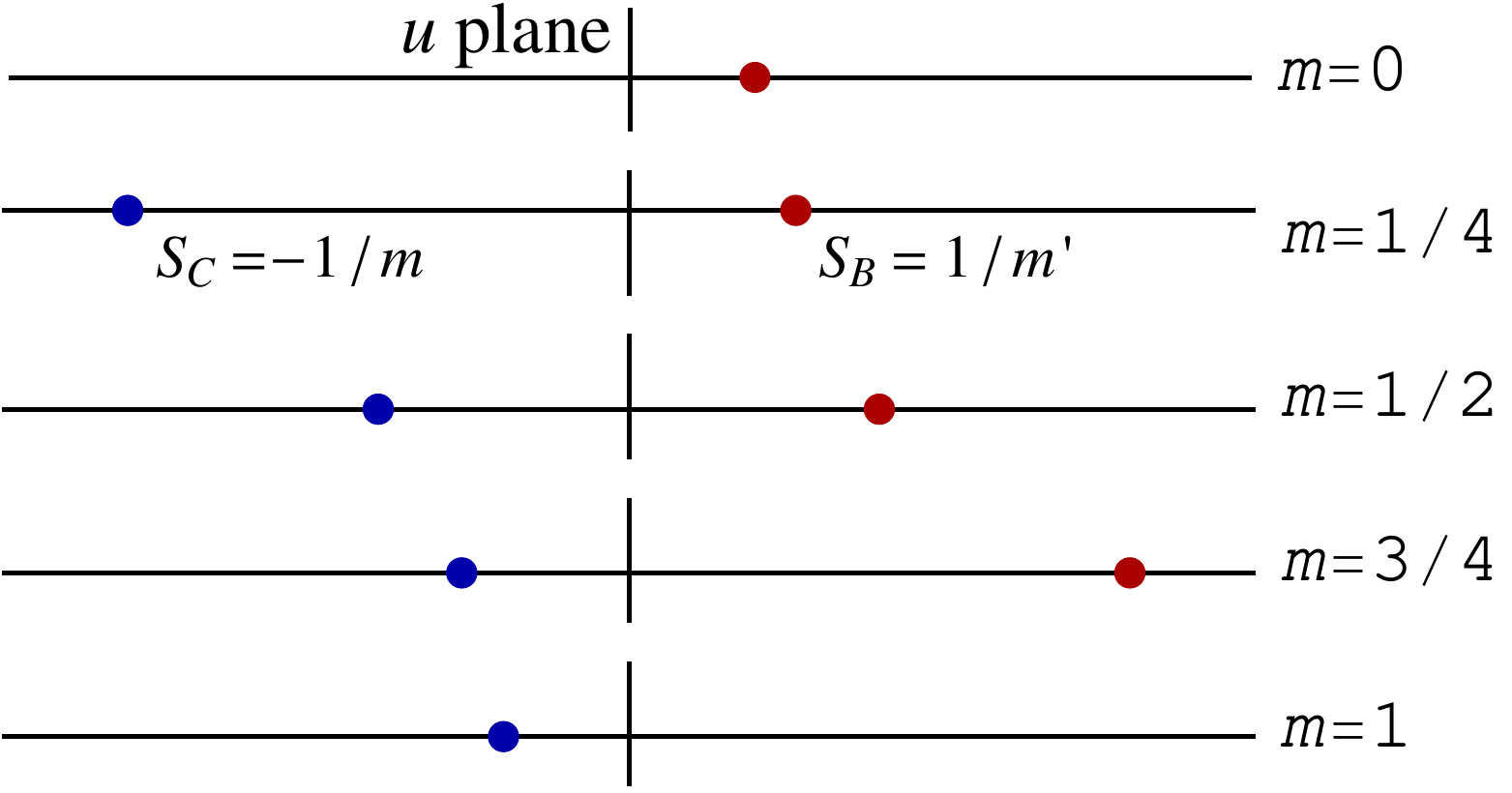} 
\caption{The complex Borel $u$-plane structure for various values of $m$ for the zero-dimensional prototype. The circles denote branch cut singularities. The singularities on $\mbb{R}^+$  and $\mbb{R}^-$  are  due to  $B$ (real) and $C$ (purely imaginary) saddles. For $0<m<1$,  $\arg(g^2) =\theta=0$ and $\theta=\pi$ direction in the coupling constant plane are Stokes lines. 
} 
\label{Fig.Borel0d}
\end{figure}

For small $g^2$, the partition function (\ref{part})  defined over cycle (\ref{cycle-C}) maps into an integral in the Borel plane with the 
 change of variables, $u= \sd^2(z |m)$,   and using the identities $\text{nd}^2(z|m)=1+m\,\sd^2(z|m)$ and  $\text{cd}^2(z|m)=1- \mpr \,\sd^2(z|m)$:
\bea
\mc I^{(A)}\left(\frac{1}{g^2} | m\right)={1\over\sqrt{\pi}\, g}\int_0^{{\infty}}du{e^{-u/g^2}\over\sqrt{u(1- \mpr u)(1+m u)}}\,.
\label{u_int}
\ea
The beauty of this transformation is that  the  action of the non-perturbative saddles are mapped to
the branch points in the Borel plane,  see Fig \ref{Fig.Borel0d} (left).\footnote{If this can be achieved in quantum mechanics or QFT, it may in principle provide an alternative definition. We thank Philip Argyres for discussions at this point.  See also \cite{Balian:1978et}} The saddle point $C$ on the imaginary axis ($z=i\Kp$) leads to the branch point $u = -\frac{1}{m}$ on the negative axis ($u\in \mbb R^-$), 
 and the saddle point $B$ on the real axis ($z=\K$) leads to the branch point $u  = \frac{1}{\mpr}$ on the positive axis ($u\in \mbb R^+$).   
 The location of these singularities (not their physical origin)  mimic the behavior of IR and UV renormalons, respectively, in asymptotically free quantum field theories.  
A generalized  partition function associated with a general cycle ${\mathfrak C}= \sum_i n_i {\cal J}_i$  is   therefore a three-term  trans-series 
(or  ``resurgent symbol'')  
with three  parameters  $\sigma_A,  \sigma_B, 
 \sigma_C$:  
\bea
\hskip -0.5 cm{\mc Z}_{{\mathfrak C}} (g^2 | m)&\equiv& 
\sigma_A \,   \Phi_A(g^2)  + \sigma_B\, e^{ - S_B/g^2 }    \Phi_B(g^2) +  \sigma_C \, e^{ - S_C/g^2}    \Phi_C (g^2) \nn\nn
&=& \sigma_A \,   \sum_{n=0}^{\infty} a^{(A)}_n(m) g^{2n} +  
\sigma_B\, e^{ - \frac{1}{\mpr g^2}}    \sum_{n=0}^{\infty} a^{(B)}_n(m) g^{2n}
 + \sigma_C \, e^{ \frac{1}{m  g^2}}    \sum_{n=0}^{\infty} a^{(C)}_n(m) g^{2n}     \qquad  
\label{stokes1}
\ea
Physically,  for $\arg(g^2) = 0$,  these three terms correspond to the vacuum sector ($A$),  and non-perturbative  real ($B$, exponentially decaying)  and complex saddles ($C$, exponentially growing) where  $\Phi_i(g^2)$ is formal asymptotic series around the given sector. 
Along certain angular directions called Stokes lines, the transseries parameters may exhibit jumps. This is called the Stokes phenomenon \cite{dingle,Berry,benderbook},  for an example, see Section \ref{sec:digress}.
In different angular sectors for $\arg(g^2)$,    
different  saddles  $\{z_i\}$  dominate the expression.  For $\arg(g^2)=0$, we can \textit{define} a partition function on a restricted cycle ${\mathcal C}_m$  (\ref{cycle-C}) so that $\sigma_C$=0 and there is no exponentially growing term. This partition function is indeed (\ref{part}). 

We now show that  the  perturbative expansions around the two non-perturbative saddles $B$ and $C$ can be combined to give 
the perturbative expansion around the perturbative vacuum,  saddle $A$. Conversely,   the vacuum (saddle $A$) perturbative expansion can be decomposed into two sub-series that can be constructed from the coefficients  of the perturbative  expansions around non-perturbative saddles $B$ and $C$. 
This is an exact realization of resurgence which asserts that perturbative expansion around distinct saddles are actually related in a precise way, as explained below. 
Expanding (\ref{u_int}) in $g^2$ we obtain the perturbative expansion (\ref{eq:zero-pert}): 
\bea
\mc I^{(A)}\left(\frac{1}{g^2} | m\right)=\sum_{n=0}^\infty a^{(A)}_n(m)\,g^{2n}=1+{g^2\over2}(1-2m)+{g^4\over8}(3-8m+8m^2)+\dots
\label{power_series}
\ea
where the coefficients in closed form are:
\bea
a^{(A)}_n(m)&=&{\Gamma(n+{1\over2})\over\pi^{3\over2}}\sum_{k=0}^n{\Gamma\left(k+{1\over2}\right) \Gamma\left(n-k+{1\over2}\right)\over\Gamma(k+1)\Gamma(n-k+1)}(-1)^k m^k m^{\prime\, n-k} \nn
&=&{\Gamma^2\left(n+{1\over2}\right)\over\pi \Gamma(n+1)}\,_2F_1\left(-n,-n,{1\over2}-n;m\right)
\label{eq:zero-closed}
\ea
Note the self duality (\ref{self_duality}) property: $a^{(A)}_n(m)=(-1)^n a^{(A)}_n(\mpr)$, which transforms the non-alternating series for $m<1/2$ to an alternating one for $m>1/2$, and the odd-index coefficients vanish when $m=\frac{1}{2}$.
 
It was proven by a well-controlled contour deformation argument in \cite{dingle,Berry}  that the perturbative expansion around any given saddle which is connected to other saddles by a steepest descent path 
(at some value of $\arg(g^2)$)  will be influenced by these ``adjacent saddles'' even when the integration cycle does not pass through 
the given adjacent saddle for a given value of $\arg(g^2)$. 
The connection is given by the {\it exact} resurgence formula:
\bea
\hskip-0.2cm \mc I^{(A)}\left(\frac{1}{g^2} | m\right)={2\over 2\pi i}\hskip -.1cm \sum_{k\in \{B,C\}} \hskip -.1cm \int_0^\infty {dv\over v}{1\over1-g^2\,v}\mc I^{(k)}(v |m)
\label{resurgence}
\ea
where the overall factor of 2 arises because the saddles adjacent to $z=0$ are $z=\pm\K,\pm i\Kp$. 
This remarkable resurgence formula states that there is an {\it exact} relation between the perturbative expansion (about saddle $A$) and the fluctuation expansions about the other non-perturbative saddles, $B$ and $C$. At leading order, this gives the corrected large-order behavior in 
(\ref{eq:d0-leading1}).

To see how this works {\it to all orders} in our  case, note that 
similar to (\ref{u_int}), the integrals $\mc I^{(B)}(m,v)$ and  $\mc I^{(C)}(m,v)$ can be put into a Borel form as well by defining $u=$sd$^2(z|m)-1/\mpr$ for the $B$ integral, and $u=$sd$^2(z|m)+1/m$ for the $C$ integral in (\ref{saddle_z}). The shifts ensure that $u=0$ on the saddle.
\begin{align}
\mc I^{(B)}(v |m)&=i\,\sqrt{{v\over\pi}}\int_0^{{\infty}}{e^{-v\,(u+{1\over \mpr})} du \over\sqrt{u(1+ m\,\mpr u)(1+\mpr u)}}\quad,\quad \cr
\mc I^{(C)}(v |m )&=i\,\sqrt{{v\over\pi}}\int_0^{{\infty}}{e^{-v\,(u-{1\over m})}du\over\sqrt{u(1- m\,\mpr u)(1-m u)}}
\label{bc_integrals}
\end{align} 
Again, the beauty of this transformation is that     the  ``relative  actions"\footnote{Relative actions are called  ``singulants" by Dingle \cite{dingle, Berry}.}  are now mapped to
the branch points in the associated Borel plane:
\begin{align}
&{\rm for \;  } B   :  \qquad \Delta S_{AB}= S_A - S_B =  -\frac{1}{\mpr},  \qquad  \Delta S_{CB}  = S_C - S_B =  -\frac{1}{m \mpr},    \cr
&{\rm for \;  } C   :  \qquad    \Delta S_{AC} =S_A - S_C =\;  +\frac{1}{m},  \qquad   \Delta S_{BC}= S_B - S_C =+\frac{1}{m \mpr},  \cr
&{\rm for \;  } A   :  \qquad    \Delta S_{BA} =S_B - S_A =\;  +\frac{1}{\mpr},  \qquad   \Delta S_{CA}= S_C - S_A =-\frac{1}{m}, 
\label{bc_relative}
\end{align} 
i.e., the Borel transform  of $\Phi_B$ has two singularities on $\mathbb R^{-}$,  while that of  $\Phi_C$ has two singularities on $\mathbb R^{+}$, and that of $\Phi_A$ has two singularities, one  on $\mathbb R^{+}$, and one on $\mathbb R^{-}$. 
 Note that   the singularities seen by the  Borel transforms of  formal power series $\Phi_i$  are therefore intimately related and  the divergences of $\Phi_i$ is controlled by $S_j - S_i$.

Expanding $(1-g^2 v)^{-1}$ in (\ref{resurgence}) in $g^2$ and using (\ref{bc_integrals}), the $v$ integrals can be carried out, which
allows us to write each coefficient of the vacuum perturbative series (\ref{power_series}) as a sum of two terms, each of which are completely determined by the saddle point integrals (\ref{bc_integrals}) around $B$ and $C$:
\bea
a^{(A)}_n(m)&=& \frac{ \Gamma(n+{1\over2})}{\pi^{3\over2}}\int_0^\infty {du\over \pi}\left({m^{\prime\,n+1}\over\sqrt{u(1+m\,\mpr u)}(1+\mpr u)^{n+1}}
+{(-1)^{n+1/2}m^{n+1}\over\sqrt{u(1-m\,\mpr u)}(1-m u)^{n+1}}\right) \nn
&=& \frac{ \Gamma(n+1)}{\pi \,(n+{1\over2})}\,_2F_1\left({1\over2},{1\over2},{3\over2}+n;1-m\right)
+(-1)^n\,\frac{ \Gamma(n+1)}{\pi \,(n+{1\over2})}\, \,_2F_1\left({1\over2},{1\over2},{3\over2}+n;m\right)
\label{resABC}
\ea
 
This is the explicit realization of resurgence in zero dimension. Perturbation theory (the expansion around the saddle $A$) is explicitly controlled, \textit{to all orders}, by the perturbative expansion around the non-perturbative saddles, $B$ and $C$, which contribute the first (non-alternating) and second (alternating) terms in  (\ref{resABC}), respectively. 
For $m<1/2$  ($>1/2$), the first  non-alternating (second alternating) series dominates. Thus, taking into account both saddles $B$ and $C$, we obtain the corrected expression (\ref{eq:d0-leading1}) for the leading large-order growth of the perturbative expansion coefficients:
\bea
a^{(A)}_n(m)&\sim  & \frac{(n-1)!}{\pi} \left[(1-m)^{n+1/2}+ (-1)^n m^{n+1/2}
\right]  
=  \frac{(n-1)!}{\pi} \left[\frac{1}{(S_B)^{n+1/2}}+  \frac{(-1)^n}{(|S_C|)^{n+1/2}}
\right]\nn\label{eq:d0-leading}
\ea
As mentioned before, this corrected leading order growth respects the self-duality property (\ref{eq:selfdual}).
 
\subsection {Higher orders of resurgent relations}

We can do even better than (\ref{eq:d0-leading}).
The {\it exact} resurgent relation (\ref{resurgence}) between saddles has a further deeper consequence:  
 {\it Early terms} of the perturbative expansions about non-perturbative saddles $B$ and $C$ give  sub-leading corrections to the {\it late terms} of the perturbative expansion around  the vacuum-$A$. From the perturbative expansions around saddles $B$ and $C$; 
\begin{align}
\mc I^{(B)}(\frac{1}{g^2} | m)&=i\,e^{-\frac{1}{\mpr\, g^2}}\,\sqrt{\mpr}\left(1-\frac{2-\mpr}{4} \mpr g^2 +\frac{3 (8-8\mpr+3m^{\prime\,2})}{32}m^{\prime\,2} g^4   -\dots\right)  \cr
&\equiv i\,e^{-\frac{1}{\mpr\, g^2}}\sum_{n=0}^\infty a^{(B)}_n(m) \,g^{2n} \nn
\mc I^{(C)}(\frac{1}{g^2} | m) &= i e^{\frac{1}{m\, g^2}}\,\sqrt{m}\left(1+\frac{2-m}{4} m\, g^2 +\frac{3 (8-8m+3m^2)}{32}m^2 g^4   -\dots\right) \cr
&\equiv i\,e^{\frac{1}{m\, g^2}}\sum_{n=0}^\infty a^{(C)}_n(m) \,g^{2n} 
\label{eq:bc_expansions}
\end{align} 
we deduce an even more accurate prediction for the large-order growth of the perturbative expansion coefficients in (\ref{eq:zero-pert}),
including the first two subleading corrections: 
\bea
a^{(A)}_n(m)&\sim  & \frac{(n-1)!}{\pi S_B^{\,n} }  \left( a_0^{(B)} + \frac{a_1^{(B)}  S_B }{(n-1)}   + \frac{a_2^{(B)}  S_B^{\,2} }{(n-1)(n-2)}    + \dots \right)  \nonumber\\
&+ &  \frac{(n-1)! }{\pi S_C^{\,n} }  \left( a_0^{(C)} + \frac{a_1^{(C)}  S_C }{(n-1)}   + \frac{a_2^{(C)}  S_C^{\,2} }{(n-1)(n-2)}    -  \dots \right)  \nonumber
\cr
&& \cr  
&& \cr  
&= & \frac{(n-1)!}{\pi} \left[m^{\prime\,n+1/2}\left(1-\frac{2-\mpr}{4(n-1)}+ \frac{3 (8-8\mpr+3m^{\prime\,2})}{32 (n-1)(n-2)}  -\dots\right) \right.\nonumber\\
&& \hskip 1cm \left. + (-1)^n m^{n+1/2}\left(1-\frac{2-m}{4(n-1)}+\frac{3 (8-8m+3 m^2)}{32 (n-1)(n-2)}  -\dots\right)
\right]
\label{eq:d0-large-order}
\ea
\begin{figure}[htb]
\begin{center}
\hspace{-0.22cm}
\includegraphics[width=25em]{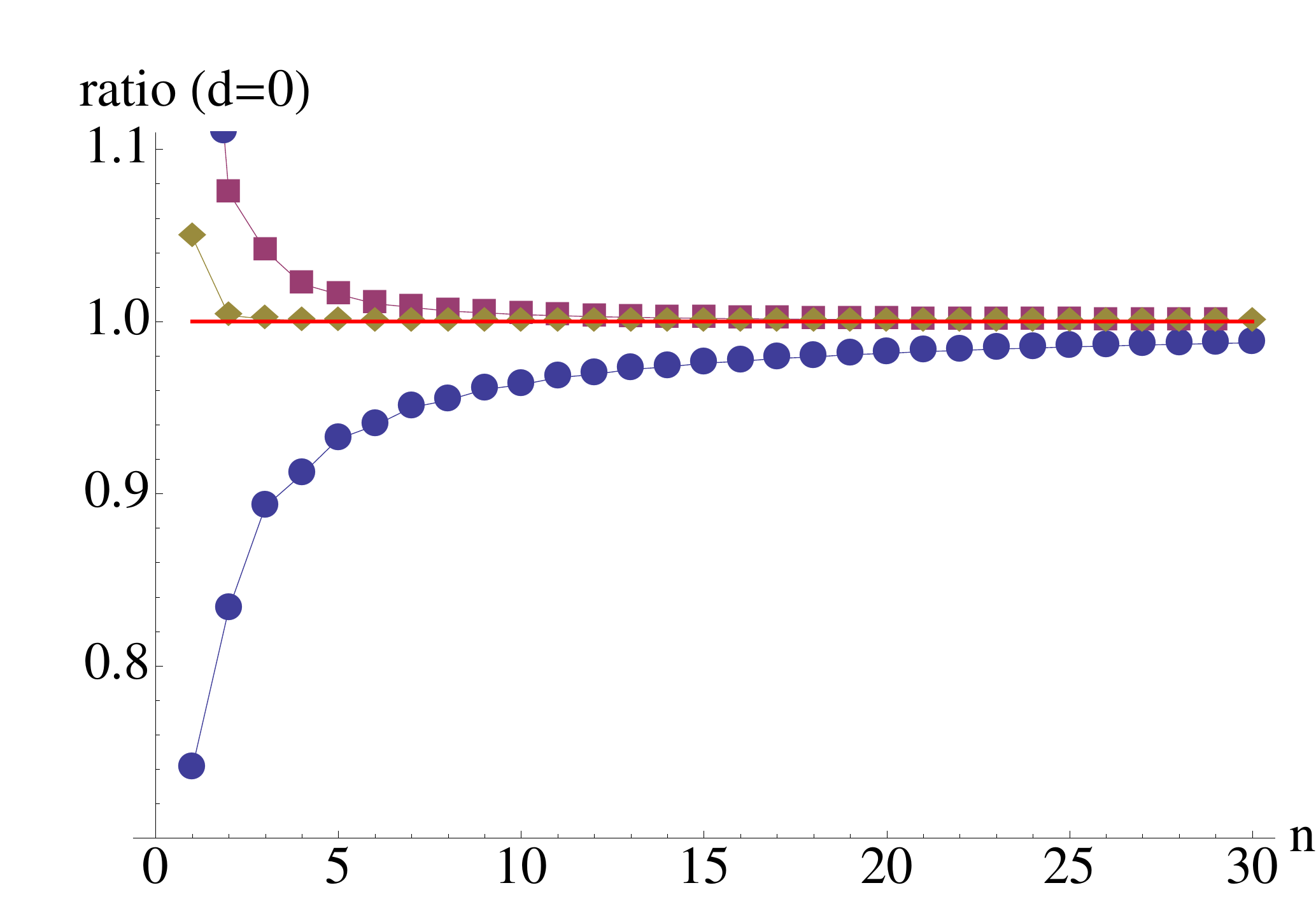}   \\
\end{center}
\caption{
The ratio ${a_n^{\rm actual}(m)}/{ a_n^{(A)}(m)} $, including the contributions to $ a_n^{(A)}(m)$ from {\it both} the  real and complex saddles, ($B$ and $C$) beyond leading order, as in (\ref{eq:d0-large-order}). These plots are all for $m=0.51$ (those for $m=0.49$ are indistinguishable)
 to leading (blue circles), sub-leading (red squares) and sub-sub leading (gold diamonds) order. 
At the sub-sub-leading order the agreement is almost perfect even for the low order terms.
} 
\label{subleading}
\end{figure}
This result is plotted in Figure \ref{subleading}, for values of $m$ close to $m=0.5$, including the leading, subleading, and sub-subleading contributions. Note the dramatic improvement coming from including the effect of the fluctuations about the $B$ and $C$ saddles. Moreover,  we observe that the sub-leading terms in (\ref{eq:d0-large-order}) respect the self-duality property (\ref{eq:selfdual}). 
When the action of real and imaginary saddles are equal in modulus, (\ref{eq:d0-large-order})  is in fact same as the one obtained earlier in 
Section 4 of  Ref.~\cite{Aniceto:2011nu}.

\begin{figure}[htb]
\center
\includegraphics[scale=0.55]{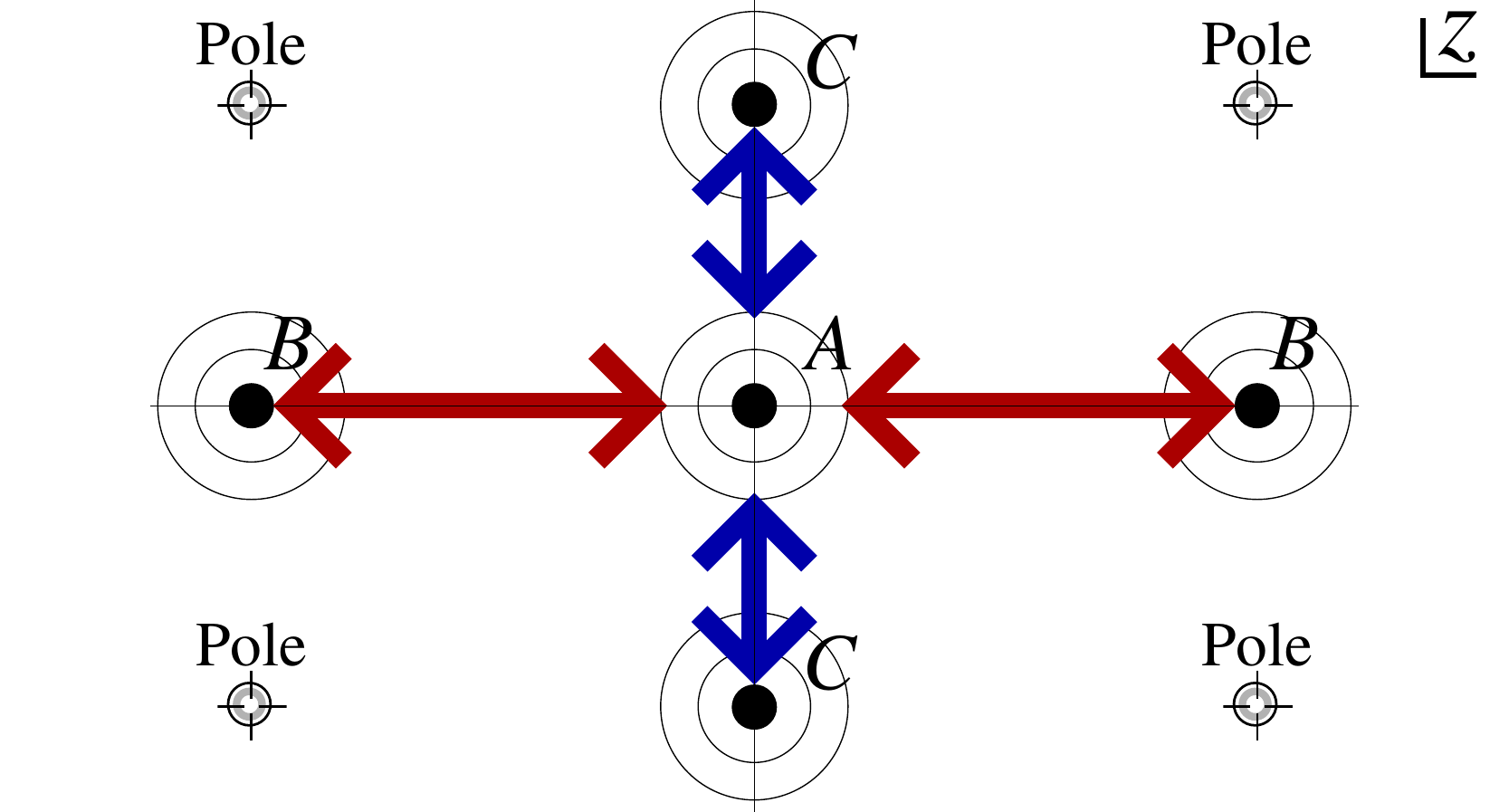}
\caption{Schematic representation of the resurgent relation (\ref{eq:d0-large-order}) between the large orders of perturbative fluctuations about the vacuum saddle $A$ given in (\ref{power_series}, \ref{eq:zero-closed})  (represented as the large circle), and low orders of fluctuations about the instanton and ghost instanton saddles $B$ and $C$ given in (\ref{eq:bc_expansions})
(represented as the small circles), respectively.}
\label{fig:darboux}
\end{figure}

Since we actually have a closed-form expression (\ref{eq:zero-closed}) for the perturbative coefficients we can also confirm {\it analytically} that this agrees with the sub-leading behavior. 
Notice the explicit correspondence of the coefficients of the sub-leading terms in (\ref{eq:d0-large-order}) and the low order terms in (\ref{eq:bc_expansions}):
\bea
a^{(A)}_n(m) =\sum_{j=0} \frac{(n-j-1)!}{\pi}\,
\left({a^{(B)}_j(m)\over S_B^{\,n-j}}+{a^{(C)}_j(m)\over S_C^{\,n-j}}\right)
\label{eq:d0-large-order-2}
\ea
This remarkable relation between high orders of fluctuations about one saddle and low orders of fluctuations about the other adjacent saddles is captured to \textit{all orders} by the exact resurgence integral relation (\ref{resurgence}) and in particular (\ref{resABC}), and is sketched schematically in Figure \ref{fig:darboux}. The above expression (\ref{eq:d0-large-order-2}) is an approximation to the exact resurgence relation (\ref{resABC}) for large $n$.  

There is an alternative way of seeing the origin of the three-term resurgent trans-series structure. 
As a function of the variable $\xi=\frac{1}{g^2}$, the partition function $\mc Z$ satisfies the following third order ODE:\bea
2m(1-m) \xi \hat{\mc Z}^{\prime\prime\prime}-\left(2(1-2m)\xi-3m(1-m)\right) \hat{\mc Z}^{\prime\prime}-2\left(\xi+(1-2m)\right) \hat{\mc Z}^\prime-\hat{\mc Z}=0
\label{eq:third-order}
\ea
where $\hat{\mc Z}=\xi^{-1/2} \mc Z$, and by a general result \cite{dingle,Marino:2012zq}, this means that $\mc Z$ has a three-term resurgent trans-series  in the large $\xi$  [small $g^2$] limit, given in (\ref{stokes1}).  
For $m=0$ and $m=1$, it reduces to a second order equation,
\bea
m=0&:&\quad \xi \hat{\mc Z}^{\prime\prime}+(\xi+1)\, \hat{\mc Z}^\prime+\frac{1}{2}\,\hat{\mc Z}=0 \\
m=1&:&\quad \xi \hat{\mc Z}^{\prime\prime}-(\xi-1)\, \hat{\mc Z}^\prime-\frac{1}{2}\,\hat{\mc Z}=0
\label{eq:second-order}
\ea
and  the partition function is a modified Bessel function, with a two-term trans-series (see Section 1.6 in  \cite{Dunne:2012ae}). 
In these cases one or other of the saddles runs off  to infinity. In these limits,  the large-order behavior of perturbation theory about the vacuum is determined by just  the other saddle. 

\subsection{Analytic continuation cycle  vs. dual cycle: role of  Lefschetz thimbles} 
\label{sec:digress}

We briefly discuss the relation between  the (unique) analytic continuation of the $d=0$ partition function  and the other  possible choices of the integration contours. This discussion is  a warm-up to the quantum phase transition discussion in the $d=1$ QM model in Section \ref{QPT}, and can also be viewed as its dimensional reduction to $d=0$ dimension.

Let us denote $\arg(g^2)\equiv\theta$. The analytic continuation of the 
partition function (\ref{part})  is given by  integration along the  contours
\bea 
\label{lt-0}
{\cal C} (\theta|m)  = \left\{ \begin{array}{lc} 
   {\cal J}_A(\theta|m) + {\cal J}_B (\theta|m)    \qquad& -\pi <  \theta< 0 \cr
     {\cal J}_A(\theta|m) - {\cal J}_B (\theta|m)  \qquad  & 0 < \theta < \pi  \cr
   - {\cal J}_A(\theta|m) - {\cal J}_B (\theta|m)  \qquad  & \pi< \theta < 2\pi  
\end{array} \right. 
\ea
The flip of the  sign of ${\cal J}_B$  at $\theta=0$  and  ${\cal J}_A$  at $\theta=\pi$  are 
examples of  the Stokes phenomenon \cite{dingle,Berry,benderbook}. Consequently, the transseries expansion for the partition function (\ref{part}) in different Stokes chambers is given by
\bea 
 {\mc Z} (g^2 | m)=  \left\{ \begin{array}{lc} 
   \Phi_A (g^2)  +  i  e^{ - \frac{1}{\mpr g^2}}  {\Phi}_B (g^2)      \qquad& -\pi <  \theta< 0 \cr
    \Phi_A (g^2)  -  i  e^{ - \frac{1}{\mpr g^2}}  {\Phi}_B (g^2)     \qquad  & 0 < \theta < \pi  
\end{array} \right. 
\label{a-c}
\ea
with a Stokes jump at   $\arg(g^2)=0$.   We first describe the physical meaning of this formal expression. 
 
 First, for $\arg(g^2) =0$, the partition function defined in (\ref{part})  is clearly real. On the other hand, 
the formal  first sum   $ \Phi_A (g^2)$ in   (\ref{a-c})    looks real, but this is deceptive. 
 Upon left/right Borel resummation  ${\cal S}_{0^\pm}$  of the formal series along $\theta= 0^{\pm}$,  we observe that  $ \Phi_A$ is non-Borel summable with a singularity at $1/m'$, while $ \Phi_B$ is  Borel summable in the $\theta= 0$ direction since its singularities are on 
 $\mathbb R^{-}$. Namely,  
   \bea 
    \Phi_A (g^2)  \pm  i  e^{ - \frac{1}{\mpr g^2}}  {\Phi}_B (g^2)   \xrightarrow{ \rm BE-summation}    && {\cal S}_{0^{\pm}}    {\Phi}_A   \pm i e^{ - \frac{1}{\mpr g^2}}    {\cal S}_{0^{\pm}}  {\Phi}_B \cr 
  = && {\rm Re} {\cal S}_{0}    {\Phi}_A  + i   \left(  {\rm Im} {\cal S}_{0^{\pm}}    {\Phi}_A    \pm  e^{ - \frac{1}{\mpr g^2}}    {\cal S}_{0}  {\Phi}_B \right)  \cr
     = && {\rm Re} {\cal S}_{0}    {\Phi}_A
\label{a-c-2}
\ea
The non-Borel summability of   $ \Phi_A$ leads to two-fold  purely imaginary ambiguity, which cancels against the jump in the 
 non-perturbative imaginary term, leading to the real (physical) result on positive real axis in coupling constant plane.
   This is an example  of Borel-\'Ecalle summability.\footnote{For quantum mechanics  and QFT  realization of Borel-\'Ecalle  summability and resurgence, see 
  \cite{Argyres:2012vv, Dunne:2012ae} and references therein. There is also an earlier discussion of resurgence in QFT literature \cite{Stingl:2002cb}. The new aspect of      
    refs.~\cite{Argyres:2012vv, Dunne:2012ae} is   the demonstration of the  generalization of the cancellation mechanism of ambiguities in quantum mechanics  \cite{Bogomolny:1980ur, ZinnJustin:1981dx, zinnbook}   to QFTs by using continuity and compactification, and also demonstration of a  semi-classical realization of infrared renormalons. 
   Other earlier work making use of resurgence theory in  string theory and large-$N$ matrix models  context (for which the renormalon divergence is absent) can be found in    \cite{Garoufalidis:2010ya,Marino:2007te,Marino:2008vx,Pasquetti:2009jg,Klemm:2010tm,Aniceto:2011nu,  Marino:2012zq,  Marino:2008ya}.}

Off the real axis,  for example, for $ 0 < \theta < \pi$, the partition function is a two-term transseries.  Evidently, in the 
$ 0 < \theta < \pi/2$ range, saddle $A$ dominates over $B$, and  $\theta= \pi/2$ is an anti-Stokes line  where the modulus of the contribution of both saddles are equal (1 vs.    $e^{  \frac{i}{\mpr |g^2|}}$) and  for  $ \pi/2 < \theta < \pi$, saddle $B$ dominates over $A$. Again, $\theta=\pi$ is a Stokes line.  

Clearly, the analytic continuation of the partition function to $\arg(g^2) = \pi^{-}$ leads to an exponential growth of the partition function, and the usual perturbative term is now the recessive term.    Despite the fact that such a partition function looks pathological at weak coupling, 
it is nonetheless the unique analytic continuation of our original partition function  (\ref{part}).\footnote{ Such exponential growth of the partition function is reported recently in the context of quantum gravity in de Sitter (dS) space, see \cite{Banerjee:2013mca}.  It would be interesting to understand the  Lefschetz thimble associated with this  growth, and if there are other homotopically independent thimbles.}  
  
 Here, we face a puzzle. Clearly, the integration cycles associated with   $[- {\cal J}_B(\pi^{-}|m) + {\cal J}_A (\pi^{-}|m)]$ and  $[- {\cal J}_B(\pi^{+})|m - {\cal J}_A (\pi^{+}|m)]$ are not the integration cycles associated with the self-duality property. Then, how is the self-duality of the  
  potential (\ref{self_duality}) realized? 
 
\begin{figure}[htb]
\includegraphics[scale=0.47]{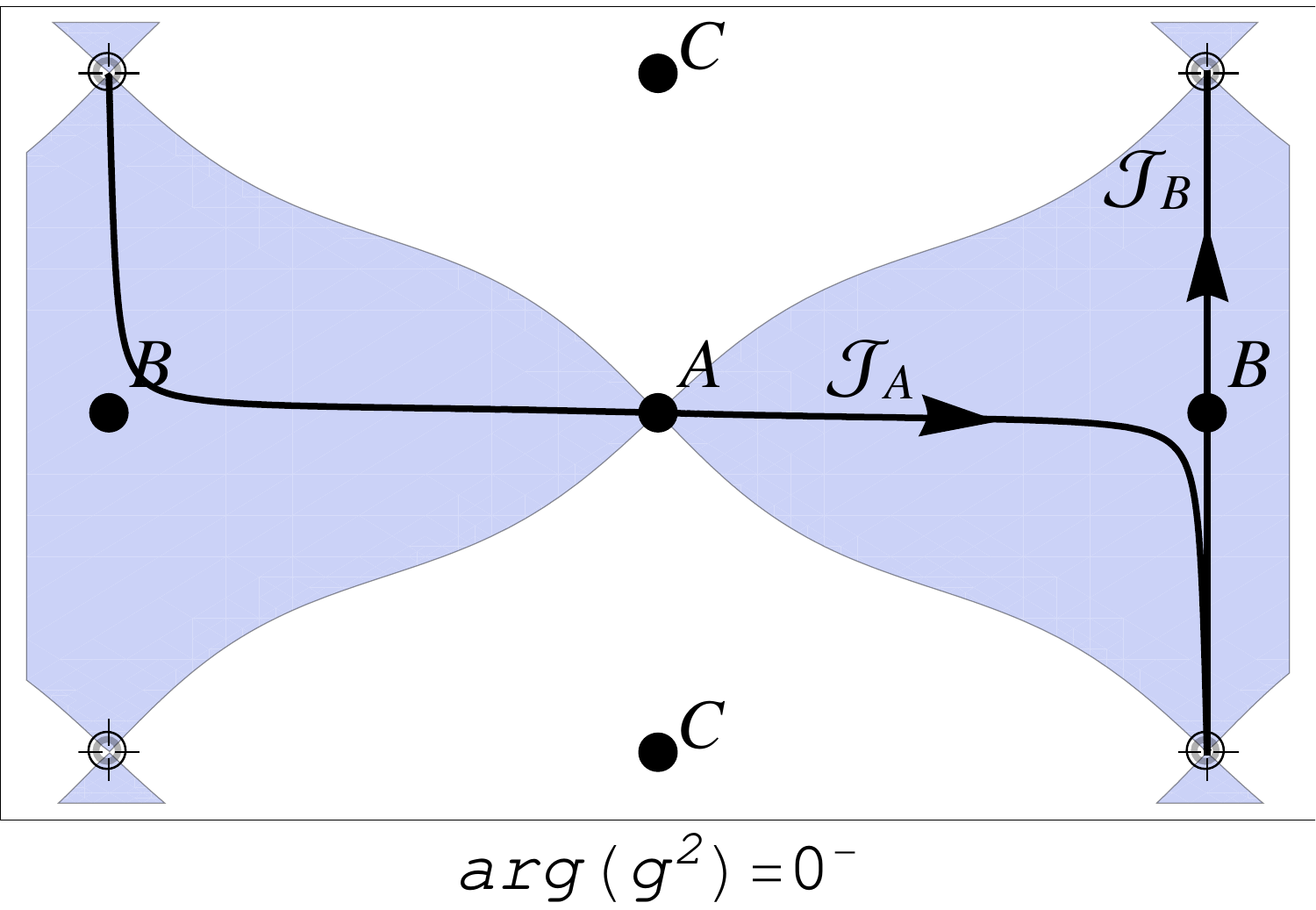}
\hspace{1cm}
\includegraphics[scale=0.47]{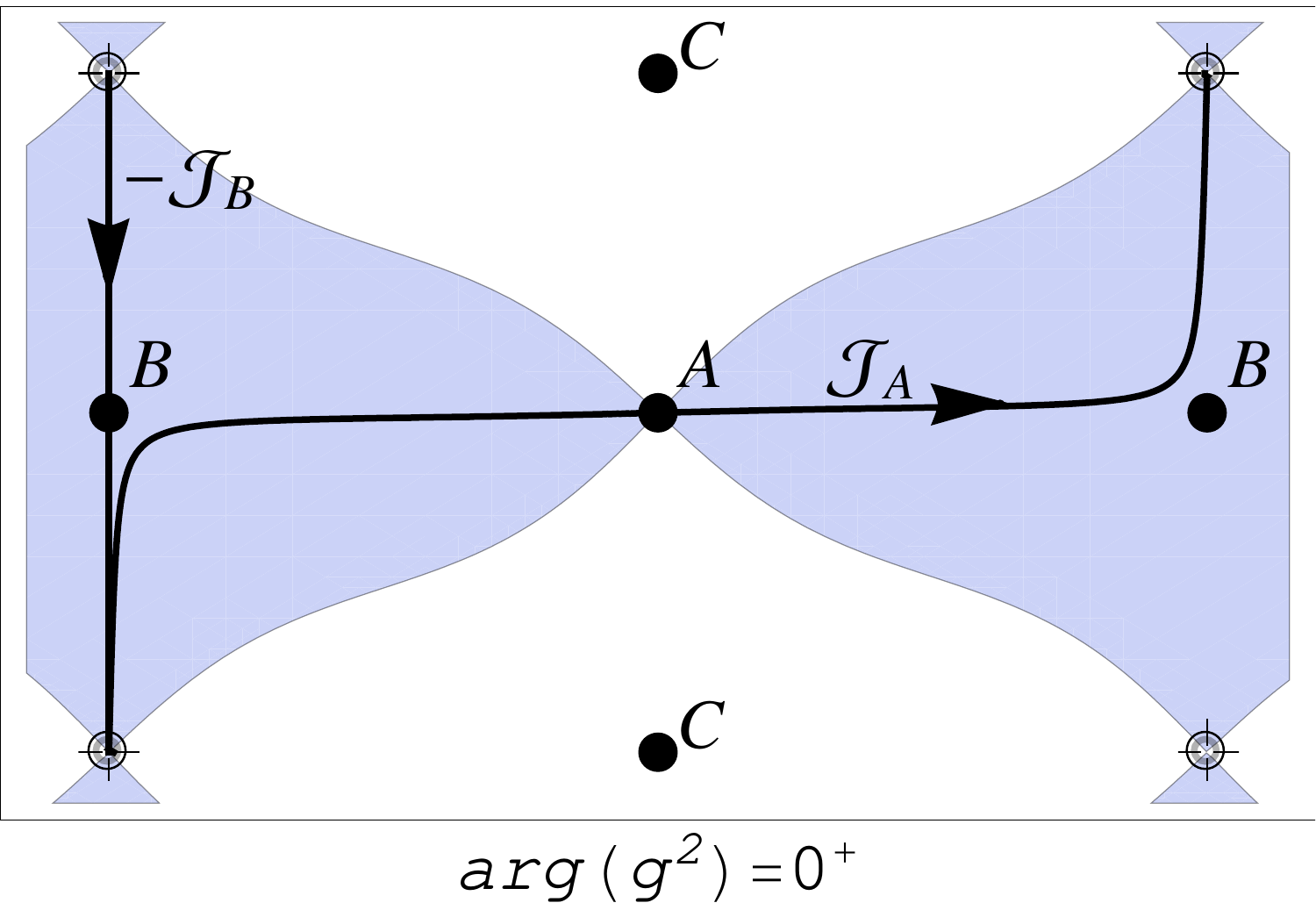}\\
\includegraphics[scale=0.47]{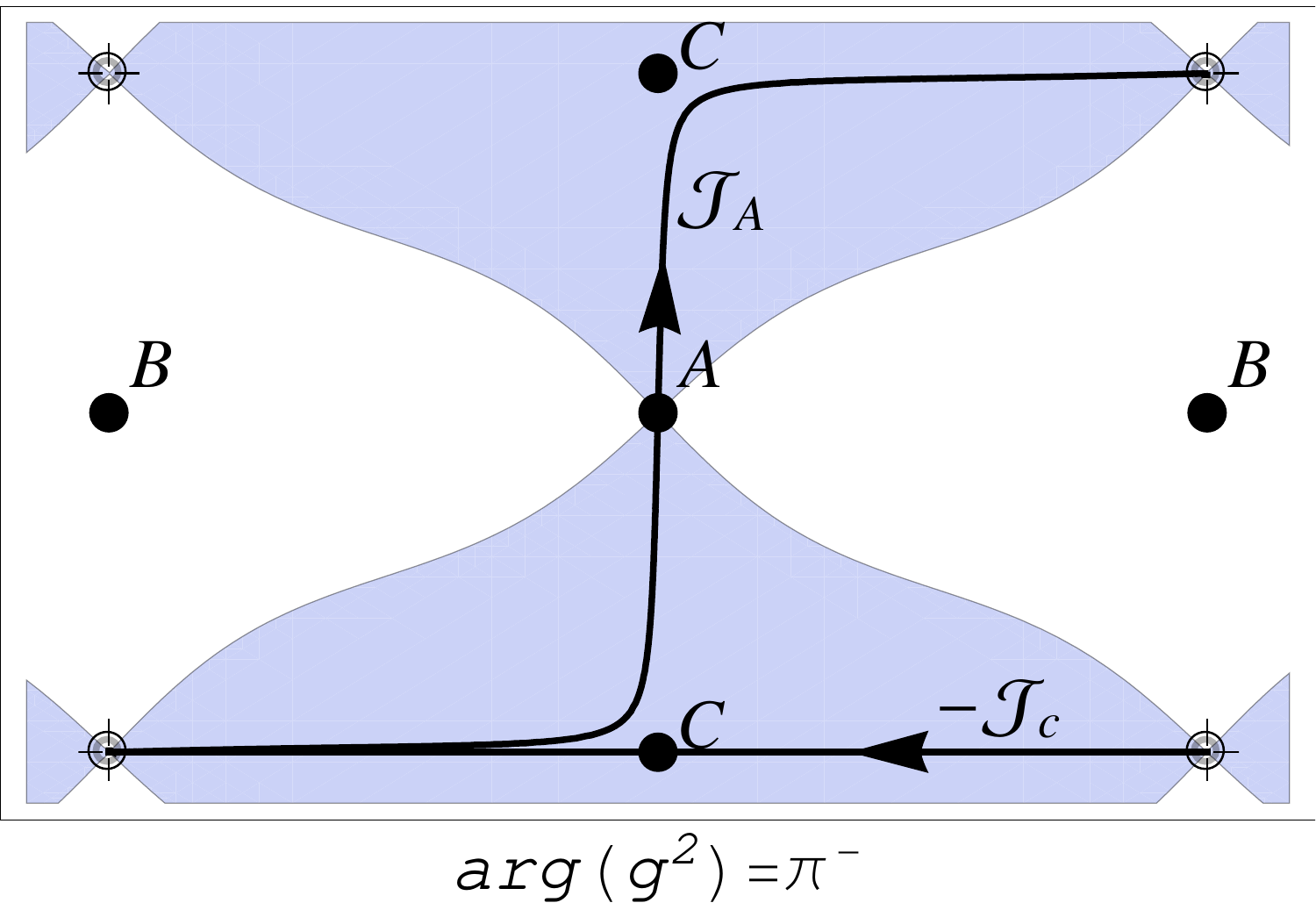}
\hspace{1cm}
\vspace{-0.5cm}
\includegraphics[scale=0.47]{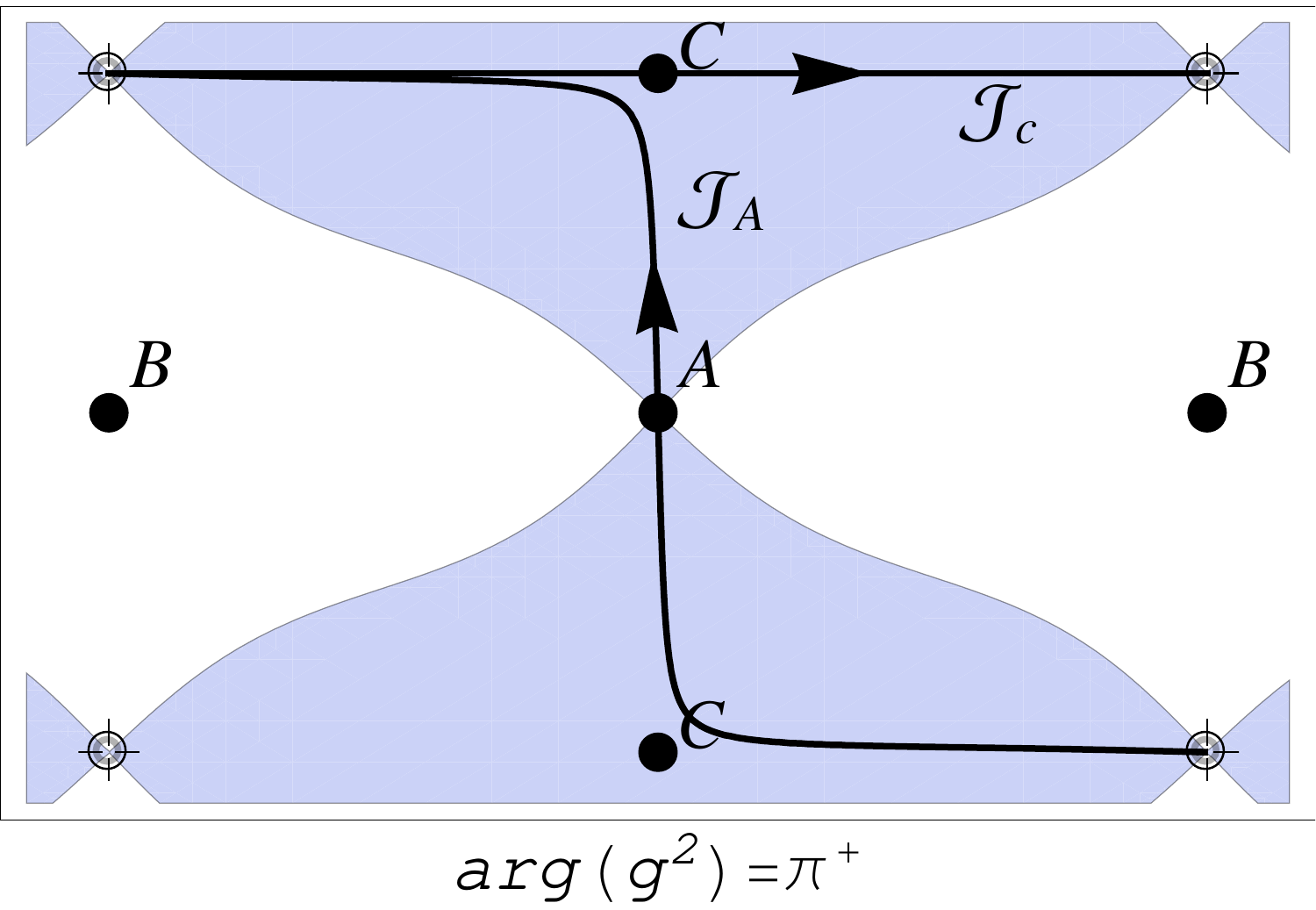}\\
\begin{center}\includegraphics[scale=0.47]{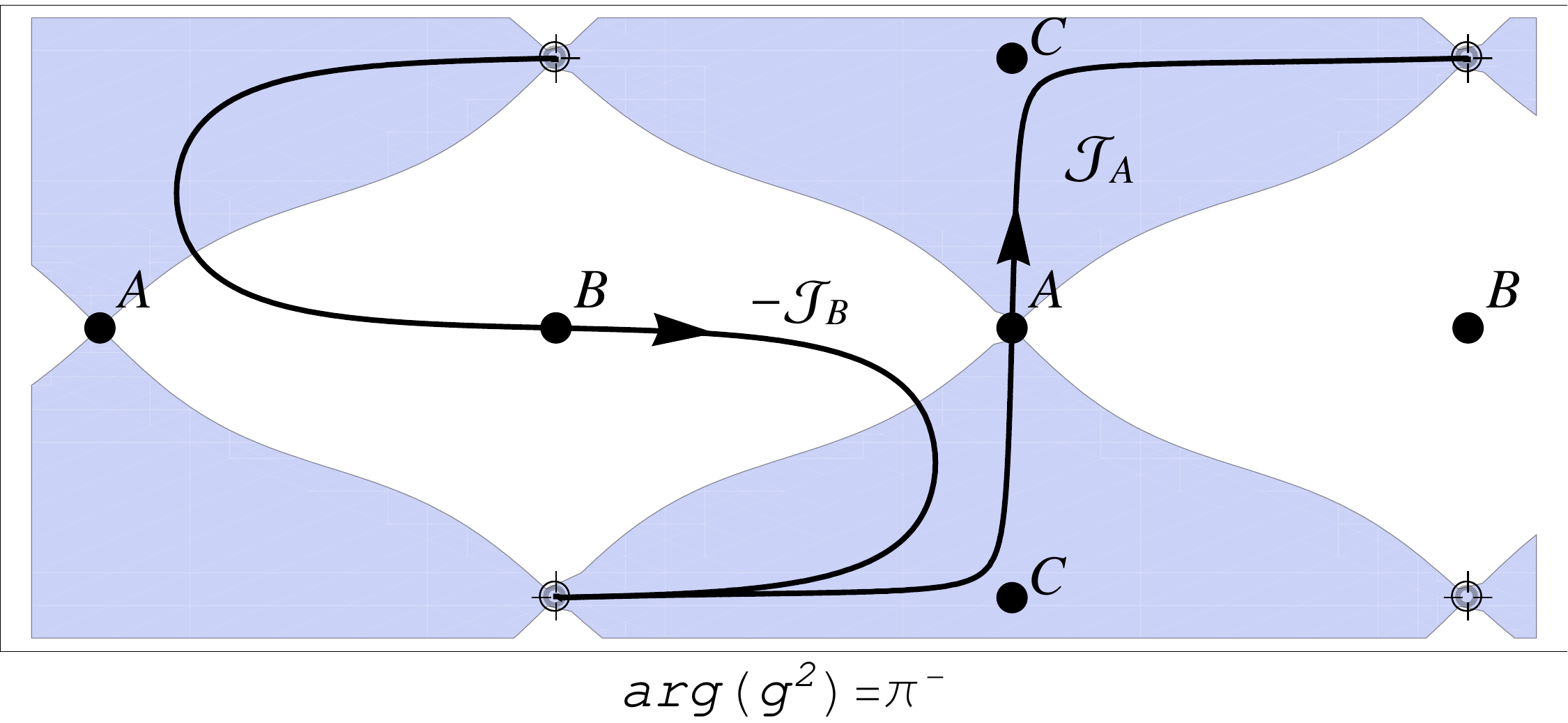}\end{center}
\vspace{-0.5cm}
\caption{The steepest descent countours (Lefschetz thimbles) shown on the fundamental torus. The blue (white) regions are where ${\rm Re}(S(z))$ is positive (negative).{\bf Upper:} The contour associated with partition function (\ref{part}) in terms of Lefschetz thimbles at $\arg(g^2)= 0^{\mp}$, 
$ {\cal C}_m= {\cal J}_A(0^{\mp}|m) \pm {\cal J}_B (0^{\mp}|m).$  
{\bf Middle:}  The contour associated with the dual cycle, $\tilde{\cal C}_m={\cal J}_A(\pi^{\mp}|m) \mp {\cal J}_C (\pi^{\mp}|m)$.
The jumps in $\cal J_B$ at $\arg(g^2)=0$ and in $\cal J_C$ at $\arg(g^2)=\pi$  are associated with Stokes phenomenon. Note that the dual cycle $\tilde{\cal C}_m$ is {\it not} an analytic continuation of ${\cal C}_m$, rather is an independent linear combination of Lefschetz thimbles.  
{\bf Lower}: The contour ${\cal J}_A(\pi^-|m) - {\cal J}_B (\pi^-|m)$ obtained from the analytical continuation of a partition function that behaves well for $g^2>0$. Note that ${\cal J}_B(\pi^-|m)$ passes through the region where ${\rm Re}(S(z))<0$ which leads to an exponentially large contribution. 
 }
\label{contours}
\end{figure}

 The resolution of this puzzle is that at any given value of $\theta$,  we in fact have three independent  integration cycles.  In Fig.~\ref{contours}, observe that at exactly $\theta=\pi$, ${\cal J}_A$ has a segment purely on imaginary axis,  
 and two segments extending parallel to real axis, 
and by using  ${\cal J}_A$ cycle in conjunction with ${\cal J}_C$, we can define an integration contour, which we call the ``dual cycle"  $\tilde {\cal C}_m$ that is purely on imaginary axis on the interval  $[- i \K', i \K' ]$.  In terms of   Lefschetz thimbles 
 \bea 
\tilde {\cal C}_m =
 \left\{ \begin{array}{l} 
   {\cal J}_A(\pi^{-}|m) - {\cal J}_C (\pi^{-}|m)  \cr
     {\cal J}_A(\pi^{+}|m) + {\cal J}_C (\pi^{+}|m) .
\end{array} \right.  \qquad 
\label{cycle-dual}
\ea
In fact, the dual cycle is related to the original cycle via a $\pi/2$ rotation of the fundamental torus and the transformation of the elliptic parameter $m \rightarrow  \mpr = 1-m$:
\bea
\tilde{\cal C}_m=i[-\Kp,\Kp]=i\,{\cal C}_{\mpr}
\ea

It is important to note that the dual cycle (\ref{cycle-dual}) is not an analytic continuation of the original cycle (\ref{cycle-C}), as explained above. In fact, as seen from Fig \ref{contours} the cycle ${\cal C}_m$ and the dual cycle $\tilde {\cal C}_{m}=i\,{\cal C}_{m^\prime}$ belong to different homotopy classes on the fundamental torus, and cannot be continuously deformed to each other as continuity would dictate.

Another way to see this discontinuity is as follows: We can define a self-dual partition function, whose associated trans-series has only $\mathcal O(1)$ and $\mathcal O(e^{-{1\over\mpr g^2}})$ terms for $g^2>0$, and only $\mathcal O(1)$ and $\mathcal O(e^{-{1\over m |g|^2}})$ terms for $g^2<0$, and no exponentially large terms in either case. This amounts to imposing $\sigma_C=0$ for $g^2>0$ and $\sigma_B=0$ for $g^2<0$, in the generalized partition function (\ref{stokes1}). This condition is very different from what the Stokes phenomenon would lead to if we analytically continued from positive to negative $g^2$. To see the difference, suppose we start with a well defined partition function for $g^2>0$ (i.e. $\sigma_C=0$ at $\theta=0$) and see whether the self duality condition is satisfied via Stokes jumps. First,  between $\theta=0^+$ and $\theta=\pi^-$ there are no jumps. Therefore there is no way that we can turn off $\sigma_B$ before passing to $\theta=\pi^+$. Second, even when we pass from $\theta=\pi^-$ to $\theta=\pi^+$, the Stokes phenomenon will only affect $\sigma_A$ and not $\sigma_B$ since $\sigma_B$ multiplies the dominant, exponentially large term at $\theta=\pi$. Consequently, $\sigma_B$ will be non-zero for $g^2<0$, thus it is impossible to satisfy the self duality condition via the Stokes jumps. We could as well have started with a well behaved function around $\theta=\pi$ (i.e. $\sigma_B=0$ at $\theta=\pi$). In this case, the analytical continuation would lead to an exponentially large term at $\theta=0$.

\section{One-dimensional case: quantum mechanics}
\label{sec:one}
We now turn to quantum mechanics, where the partition function is a path integral: 
\begin{eqnarray}
{\mathcal Z}(g^2|m)=\int {\mathcal D}\phi\, e^{- S[\phi]}  = \int {\mathcal D}\phi\, e^{-\int d\tau \left( \frac{1}{4}  
\dot \phi^2
 + \frac{1}{g^2}\, \sd^2(g\,\phi|m) \right)
  } 
\label{path-int}
\end{eqnarray}
(In our convention, we set the mass=1/2).   Our goal is to  generalize the machinery of resurgence theory and analytic continuation 
to path integrals. In particular, we aim to understand the role of generalized instanton sector \cite{Aniceto:2011nu}
(instantons and   ghost-instantons, in our case)
in physical observables, and in large orders in perturbation theory. Remarkably, we find results that mimic closely the behavior of the $d=0$ prototype model of the previous section.

\subsection{Perturbative analysis at large order}
\label{sec:d1perturbative}
 
Standard Rayleigh-Schr\"odinger perturbation theory for the ground state energy yields: 
\begin{align}
E^{(0)}(g^2 | m)&=1+{g^2\over2}\left(m-\frac{1}{2}\right)+{g^4\over2^3} \left(m^2-m-\frac{1}{2}\right)+{g^6\over 2^5} \left(3m-\frac{3}{2}\right)  \cr
&+{g^8\over2^7} \left(-m^4+2 m^3-18 m^2+17 m-\frac{53}{8}\right) \nn
&+ {3 g^{10}\over 2^{12}} \left(256 m^3-384 m^2+326 m-99\right)+\dots
\label{eq:gs}
   \end{align}
For example, for $ m=\{ 0, \frac{1}{4},  \frac{1}{2},   \frac{3}{4}, 1\}$ (compare with expressions (\ref{d0e})):
   \bea
E^{(0)}(g^2| 0)&=&1-\frac{g^2}{4}-\frac{g^4}{16}-\frac{3 g^6}{64}-\frac{53 g^8}{1024}-\frac{297 g^{10}}{4096} -\dots 
\\\nn
E^{(0)}(g^2| 1)&=& 1+\frac{g^2}{4}-\frac{g^4}{16}+\frac{3 g^6}{64}-\frac{53 g^8}{1024}-\frac{297 g^{10}}{4096}-\dots 
\\\nn
E^{(0)}\left(g^2 \bigg | \frac{1}{4}\right) &=&1-\frac{g^2}{8}-\frac{11 g^4}{128}-\frac{3 g^6}{128}-\frac{889 g^8}{32768}-\frac{225 g^{10}}{8192}-\dots 
\\\nn
E^{(0)}\left(g^2 \bigg | \frac{3}{4}\right) &=&1+\frac{g^2}{8}-\frac{11 g^4}{128}+\frac{3 g^6}{128}-\frac{889 g^8}{32768}+\frac{225 g^{10}}{8192}-\dots 
\\\nn
E^{(0)}\left(g^2 \bigg | \frac{1}{2}\right)&=&1+0g^2-\frac{3 g^4}{32}+0g^6-\frac{39 g^8}{2048}+0g^{10}
-\dots
\ea
Notice  the self duality relation: $E^{(0)}(g^2 | m)= E^{(0)}(-g^2 | 1-m)$. For $m<\frac{1}{2}$ the series is non-alternating, 
   while for $m>\frac{1}{2}$, the perturbative series is alternating. 
    When $m=\frac{1}{2}$, all odd terms vanish, and the remaining series is in  $g^4$. This behavior is precisely parallel to that found in the $d=0$ example in the Section \ref{sec:zero}.
  
In one-dimensional QM, the  potential     (\ref{pot}) has a real instanton, a solution of $\dot{x}(t)= 2\sqrt{V(x(t))}$, given by:
    \bea
    x_{\mathcal I}(t)=\frac{1}{g}\,{\rm cn}^{-1}(\tanh (-2t)|m)
    \label{eq:inst}
    \ea
    which  interpolates between  $0,2\K$ along the real axis (see the second figure in Fig \ref{fig:rectangle}). The
corresponding instanton action is:
  \bea
 {S_{\cal I}(m)\over g^2} &=&\frac{1}{g^2}\int_0^{2\K}\sd(x|m)\,dx={2\sin^{-1}(\sqrt{m})\over g^2 \,\sqrt{m \mpr}}\geq \frac{2}{g^2}  
\label{actions}
\ea
There is also a complex instanton solution, the ``ghost instanton'' ${\mathcal G}$:
\bea
x_{\mathcal G}(t)=\frac{i}{g}\,{\rm cn}^{-1}(\tanh (-2t)|\mpr)
\label{eq:ghost}
\ea
which interpolates between  $0,2i\Kp$ along the imaginary axis (see Fig \ref{fig:rectangle}). The associated 
ghost instanton action is:
  \bea
 {S_{\cal G}(m)\over g^2} &=&\frac{i}{g^2}\int_0^{2i\Kp}\sd(x|m)\,dx=- {2\sin^{-1}(\sqrt{\mpr})\over g^2\,\sqrt{m \mpr}} \leq -\frac{2}{g^2} 
\ea
Note the self-duality relation: $S_{{\mathcal G}}(m) = - S_{\cal I}(\mpr)$. The instanton (ghost-instanton) action is positive (negative) definite  for $\arg(g^2) =0$. Also note that  under $g^2 \rightarrow -g^2$, instanton solution becomes a ghost-instanton and vice-versa.

 Since the original path integration is over real paths $x(t)$, we might naively think that only the real instanton affects the dynamics and observables. 
 Then we would predict that to leading order, the large-order growth of the perturbative coefficients of the ground state energy is controlled by the instanton/anti-instanton ($[{\cal I} \bar {\cal I} ]$ for short) events:
\bea
a_n^{\rm naive}(m)\sim -\frac{16\,  n! }{\pi}\,\frac{1}{(S_{{\cal I}  \bar {\cal I}}(m))^{n+1}}
\label{eq:d1-naive}
\ea
which is the analogue of (\ref{eq:d0-naive}). Here $S_{\cal I \bar{\cal I}}(m)\equiv2\, S_{\cal I}$ is the  $[{\cal I} \bar {\cal I} ]$ action. 

We have calculated the first 30 terms in the perturbative expansion (\ref{eq:gs}). There are many ways to do this. We used the Zinn Justin/Jentschura approach \cite{zjj}, in which one expands the exact WKB quantization condition in a perturbative series in  the coupling constant, and then inverts to express the energy in terms of the level index. In fact this gives perturbative expressions for every energy level, but here we only quote the results for the ground state. We confirmed that this agrees with the result obtained  by Bender-Wu recursion  relations \cite{Bender:1969si},  matching the $m=0$ (Sine-Gordon) and $m=1$ (Sinh-Gordon) cases  analyzed by Stone and Reeve \cite{stone}.

\begin{figure}[htb]
\begin{center}
\hspace{-0.3cm}
\includegraphics[width=25em]{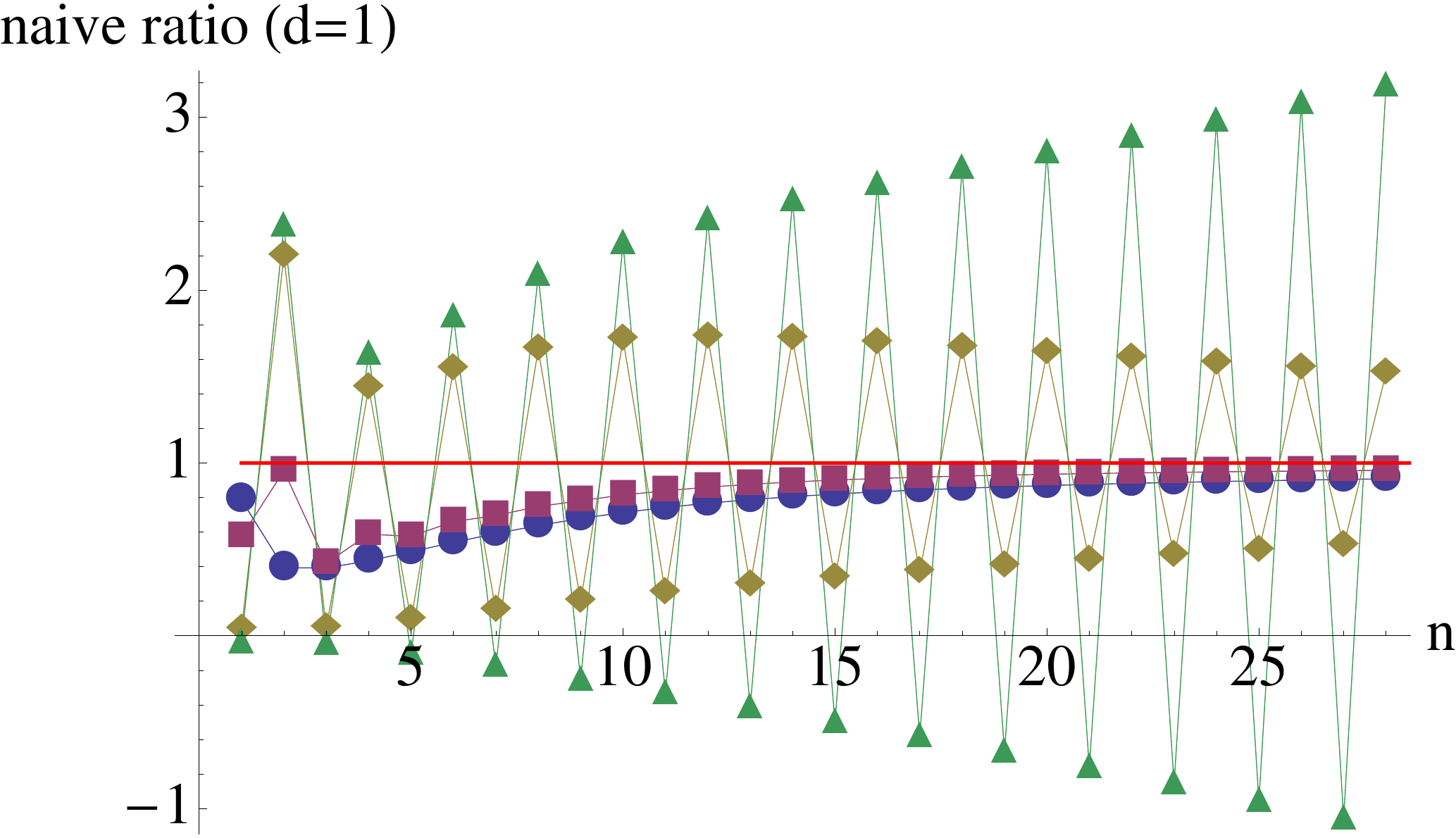}
\includegraphics[width=25em]{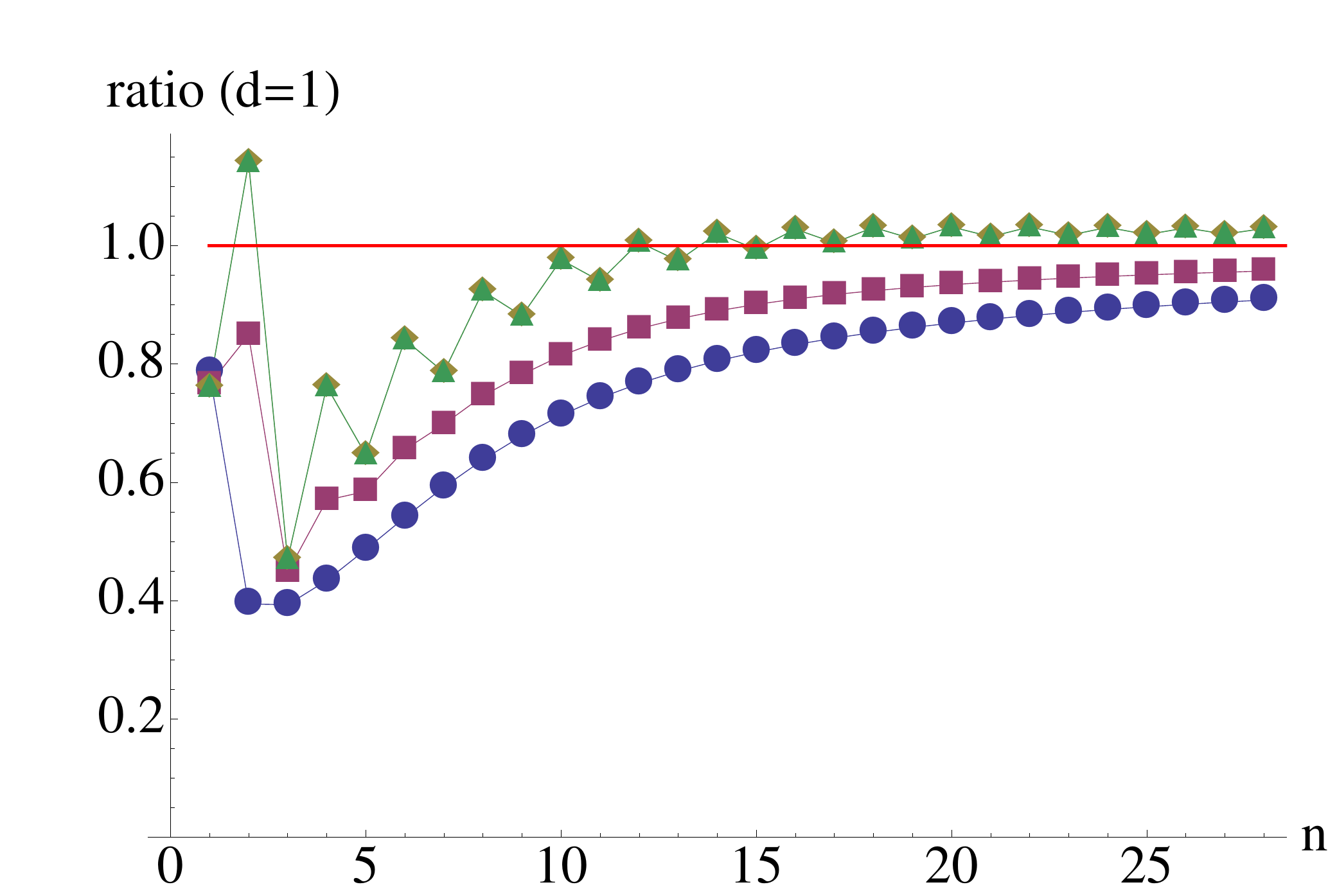} \\
\end{center}
\caption{{\bf Upper:}  ${a_n^{\rm actual}(m)}/{ a_n^{\rm naive}(m)} $: The ratio of the actual vacuum perturbation series coefficients to the ``naive'' prediction (\ref{eq:d1-naive}) for the large order growth which does not include the effect of ghost instanton events $[ {\mathcal G}  \bar {\mathcal G} ]$.
The different curves refer to different values of the elliptic parameter $m$:  $m=0$ (blue circles), 
 $m=\frac{1}{4}$ (red squares), $m=0.49$ (gold diamonds), and $m=0.51$ (green triangles).   As $m$ approaches $1/2$ from below the agreement breaks down rapidly, showing that the contribution of the $[ {\mathcal I}  \bar {\mathcal I} ]$ events by itself is not sufficient to capture the large order growth. \newline
{\bf  Lower:} The ratio ${a_n^{\rm actual}(m)}/{ a_n(m)} $, including the contributions to $ a_n(m)$ from {\it both} the $[ {\mathcal I}  \bar {\mathcal I} ]$ and $[ {\mathcal G}  \bar {\mathcal G} ]$ saddles, as in (\ref{eq:answer}). The different curves refer to different values of the elliptic parameter $m$:  $m=0$ (blue circles), 
 $m=\frac{1}{4}$ (red squares), $m=0.49$ (gold diamonds), and $m=0.51$ (green triangles). The $m=0.49$ and $m=0.51$ cases are in fact identical, a reflection of self-duality. Note the dramatically improved agreement compared to the upper figure based on (\ref{eq:d1-naive}), especially for  $m\geq 1/2$.
} 
\label{fig:d1-answers}
\end{figure}

As in the $d=0$ case, the prediction (\ref{eq:d1-naive}) is good for small $m$, but becomes bad as $m$ approaches and exceeds $\frac{1}{2}$, as shown in the first figure of Fig \ref{fig:d1-answers}. 
Again, as in $d=0$, the resolution of this problem is that there also exist complex (ghost) instantons that contribute to the the large order growth.  Taking into account the contribution from both $[ {\cal I}  \bar {\cal I} ]$ and   $ [ {\mathcal G}  \bar {\mathcal G} ]$, we are led to a corrected form of 
 leading large order growth of the perturbative coefficients:
\bea
a_n(m)\sim -\frac{16}{\pi} n!\left(\frac{1}{( S_{\cal I \bar{\cal I}}(m))^{n+1}}-\frac{(-1)^{n+1}}{| S_{\cal G \bar{\cal G}}(m)|^{n+1}}\right)
\label{eq:answer}
\ea
with a contribution from {\it both} $[ {\cal I}  \bar {\cal I} ]$ and  $[ {\mathcal G}  \bar {\mathcal G} ]$. Here  $ S_{\cal G \bar{\cal G}}(m)\equiv2\,  S_{\cal G}$ is the  $[ {\mathcal G}  \bar {\mathcal G} ]$ action. Recall the self-duality property $ S_{\cal I}(m)=- S_{\cal G}(1-m)$, which implies that the expression (\ref{eq:answer}) explicitly satisfies the self-duality condition (\ref{eq:selfdual}), unlike the naive expression  (\ref{eq:d1-naive}). \footnote{Conversely,  knowing the  naive result (\ref{eq:d1-naive}) controlled by instanton action  and 
demanding  the 
 self-duality condition (\ref{eq:selfdual})  results  in  (\ref{eq:answer}). Via this procedure, we would also predict the existence of  a ghost-instantons. Such duality-symmetric  resurgent transseries may be useful in field theories which exhibit S-duality or other dualities, see the conclusions and reference therein.}

In the second figure of Fig \ref{fig:d1-answers}, we show that with the inclusion of both $[ {\cal I}  \bar {\cal I} ]$ and  $[ {\mathcal G}  \bar {\mathcal G} ]$ terms,  the large order growth in (\ref{eq:answer})  is in excellent agreement with the actual growth, for all values of $m$, both smaller and greater than $1/2$. These numerical tests shown in Fig~\ref{fig:d1-answers} confirm that 
both instantons and ghost-instantons  do indeed control the large-order growth of perturbation theory, as characterized by the resurgent expression (\ref{eq:answer}).   Notice the similarity between the two figures in Fig \ref{fig:d1-answers} and the two figures in Fig \ref{all-fig}.

There are several limits of particular interest:
\\
${\bm m=0:}$   $S_{\cal G}(m)\to +\infty$, and the ghost-instanton sector disappears. The series is purely non-alternating for the Sine-Gordon model, and non-Borel summable with 
$a_n(0)\sim -  2^{2-n}n!/\pi$.
The leading large-order growth is  controlled solely by  $[ {\cal I}  \bar {\cal I} ]$ saddles. 
\\
${\bm m=1}$: $=S_{\cal I}(m)\to +\infty$ and the real instanton sector disappears. The series is purely alternating for the Sinh-Gordon model, and Borel-summable with
$a_n(0)\sim -  (-1)^{n}2^{2-n}n!/\pi$.
The leading large-order growth is  controlled solely by   $ [ {\mathcal G}  \bar {\mathcal G} ]$ saddles. 
\\
${\bm m=1/4}$:  The  series is  non-alternating (non-Borel summable).  It is a sum of a dominant 
non-alternating  (non-Borel summable) sub-series  controlled by  $[ {\cal I}  \bar {\cal I} ]$ events and a recessive 
alternating  (Borel summable)
sub-series controlled by  $[ {\mathcal G}  \bar {\mathcal G} ]$ saddles. 
 \\
${\bm m=3/4}$: The  series is  alternating, but this does not mean it is  Borel summable.  It is a sum of a recessive 
non-alternating series  controlled by   $[ {\cal I}  \bar {\cal I} ]$ saddles, and a dominant 
alternating series controlled by   $ [ {\mathcal G}  \bar {\mathcal G} ]$ saddles.  \\
${\bm m=1/2}$: Since the action of the real and ghost instanton  are equal in modulus ,  but opposite in sign,  the odd coefficients vanish, leaving a combined non-alternating series in $g^4$.  However, this does not imply  that there are non-perturbative effects of the form $e^{-1/g^4}$.  Rather,  it is a sum of  non-alternating and alternating  series  due to  $[ {\cal I}  \bar {\cal I} ]$  and $ [ {\mathcal G}  \bar {\mathcal G} ]$  events, respectively. Since $S[1/2]=\pi$, we have 
$a_n(\frac{1}{2})\sim - 2^{3-n} n! (1+(-1)^n )/\pi^{n+2} $.

\subsection{Resurgent trans-series and QM perturbation theory}

Let us now elaborate on the general structure of resurgence in the presence of ghost instantons.  Since we can add a topological $\Theta$ angle to a system with a periodic potential, and $\Theta$ does not appear in perturbation theory,  the set of saddles that perturbation theory around the perturbative vacuum can ``talk with"  must  be $\Theta$ independent. This is why the leading term in the large order growth of vacuum perturbation series is given by $[ {\cal I}  \bar {\cal I} ]$ and   $ [ {\mathcal G}  \bar {\mathcal G} ]$. In general, the (real) instanton/anti-instanton sector, $\{ [ {\cal I}  \bar {\cal I} ], [ {\cal I}^2  \bar {\cal I}^2 ], \ldots  \} $ may be in the same transseries as the perturbative vacuum, in accordance with the rules of the resurgence triangle described in \cite{Dunne:2012ae,du:qm}. 
Similarly, there is a resurgence triangle for  ghost-events,  and this tells us that the ghost/anti-ghost sector, 
$\{ [ {\mathcal G}  \bar {\mathcal G} ], [ {\mathcal G}^2  \bar {\mathcal G}^2 ], \ldots \} $  may also be in the same transseries as the perturbative vacuum.   

The full transseries (for the $\Theta$ independent component of the partition function, the other  topological sectors can be treated similarly)  must incorporate the exponentials associated with paired events in the long chain
 \begin{align}\label{trans-2}
\begin{matrix}
 \ldots    \longleftrightarrow    { [{\cal G}^2 \bar{\cal G}^2]}     & \longleftrightarrow &  { [{\cal G} \bar{\cal G}] }    & \longleftrightarrow & {\rm pert. vac} &  \longleftrightarrow & [{\cal I} \bar{\cal I}]   & 
      \longleftrightarrow  [{\cal I}^2 \bar{\cal I}^2]        \longleftrightarrow \ldots  \, ,  \\
\end{matrix}
\end{align}
with a transseries expansion similar to  (\ref{stokes1}), but now involving infinitely many decaying and growing exponential factors 
with their transseries parameters. 
\begin{figure}[htb]
\center
	\includegraphics[scale=0.55]{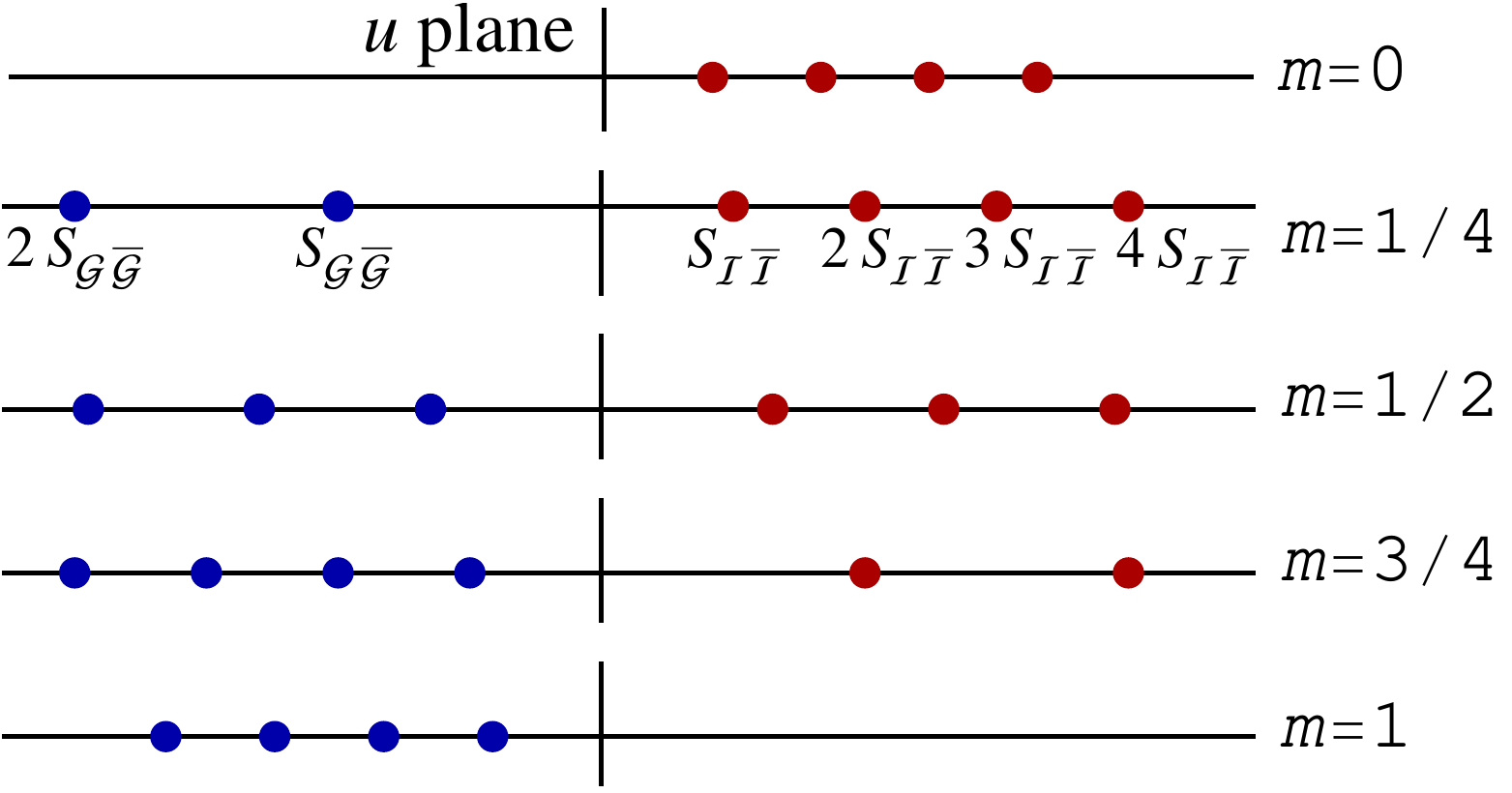}
\caption{The complex Borel $u$-plane structure for   various values of $m$ for $d=1$, quantum mechanics.   The circles denote branch cut singularities.  The leading singularities on $\mbb{R}^+$ and $\mbb{R}^-$   are  due to $[ {\cal I}  \bar {\cal I} ]$  and $[ {\mathcal G}  \bar {\mathcal G} ]$ saddles.}
\label{Fig.Borel}
\end{figure}

The paired events shown in (\ref{trans-2})  are the source of singularities in the Borel plane, located at  (see Fig \ref{Fig.Borel}):
\bea
u^{+}_k =  k [S_{\cal I\bar I}(m)]  \quad,\quad u^{-}_k= k [S_{\cal G \bar G}(m)] 
\ea 
Resurgence theory predicts  that both $u^{+}_k$ and $u^{-}_k$ control the growth of perturbation theory.

All earlier QM examples considered in the literature correspond  to cases like either  $m=0$ or $m=1$ limits, where there are either only real or  only purely imaginary  instantons.  Although this is the first  QM example that we know of with singularities on both 
$\mathbb R^{+}$ and $\mathbb R^{-}$  (mimicking  higher dimensional QFTs, e.g., QCD, ${\mathbb C}{\mathbb P}^{N-1}$),   it is actually generic to have singularities along  multiple  rays emanating from the origin  in the Borel plane for general potentials, not only $\theta=0, \pi$ directions.

\subsection{Analytic continuation cycle vs. dual cycle of path integral}
\label{sec:ac}
Defining  $x(\tau) = g \phi (\tau)$ in (\ref{path-int}), we can re-write  the partition function as  
\begin{eqnarray}
{\mathcal Z}(g^2|m)= \int_{{\Gamma} (0|m) } {\mathcal D} x \, e^{- \frac{1}{g^2} \int_0^{\beta} d\tau \left( \frac{1}{4}  
\dot x^2
 +\, \sd^2( x |m) \right)
  } 
\label{path-int-2}
\end{eqnarray}
where  we dropped an over-all normalization factor, and $ \Gamma(0|m)  $ stands for the integration cycle which is over real fields; i.e.,  $x (\tau)  \in \mathbb R$. \footnote{For a general quantum mechanical problem, a lattice regularization of the path integral  results in a multi-dimensional coupled ordinary integral, on  ${\mathbb R}^N$, where $N$ is the number of sites.  
%(In our case, if we work on a circle, then,  we are working on the space of periodic functions, hence the multi-dimensional integral can be defined on a quotient space,  $(S_1)^N =  ({\mathbb R}/{\mathbb Z})^N$).
 In order to describe analytic continuation, it is useful to view  ${\mathbb R}^N$ as embedded into ${\mathbb C}^N$ and view the partition function  as a mapping from ${\mathbb C}^N \longrightarrow {\mathbb C}$.  This complexifies  the field, but integration is  defined on a sub-manifold $\Sigma^N \in {\mathbb C}^N$, hence no redundant degrees of freedom are introduced in the integration \cite{pham, Witten:2010cx}. Also see \cite{Cristoforetti:2012su}.
 
 Similar to ordinary single  integrals discussed in the previous section, one can associate to each critical point of the action,  an integration cycle of real dimension $N$, immersed into   ${\mathbb C}^N$.  These sub-manifolds are called 
Lefschetz thimbles. To each critical point, 
\bea
t_i \in   \{ {\rm pert.   \; vacuum},  \;  {\cal I},  \bar {\cal I} ,    [ {\cal I}  \bar {\cal I} ],  \ldots,  {\mathcal G},  \bar {\mathcal G} ,  [ {\mathcal G}  \bar {\mathcal G} ], \ldots \}
 \ea
 we can associate a Lefschetz thimble ${\cal J}_i$. 
A generalized  partition function is associated with a general cycle 
\bea
{\Sigma}^N = \sum_i n_i {\cal J}_i, \qquad n_i \in \mathbb Z  
\ea
Generically, each  Lefschetz thimble ${\cal J}_i$ produce a complex number, including ${\cal J}_0$. 
 The partition function defined in 
(\ref{path-int-2}), and many other partition functions in field theories,   possibly  require   an extremely structured  linear combination of 
infinitely many ${\cal J}_i$, forcing the resurgent transseries to remain real  on positive real axis in coupling constant plane \cite{Argyres:2012vv,  Dunne:2012ae, A-S}.
Even in the simplest case on $d=0$ dimensional theory with an ordinary integral, note  that obtaining our ordinary partition function requires 
two out of three thimbles (\ref{lt-0}) and a two term transseries  (\ref{a-c-2}). In the text, we refer to the integration cycle associated with the continuum limit  as  $ \Gamma(\theta|m) \in  \mathbb C^{\infty} $. 
\label{difficult}
 }

We can in principle do an analytic continuation of the path integration to $g^2 \rightarrow g^2 e^{i \theta}$, and this will deform the integration cycle into some $\Gamma(\theta|m)$. This is in general difficult, see Footnote.\ref{difficult}.   However,  
the most interesting case for us is $g^2 \rightarrow -g^2$ continuation, for which we can develop some intuition based on $d=0$ example. The Borel plane singularity structure, as shown in Fig.~\ref{Fig.Borel}, closely mirrors that of the $d=0$ case (shown in Fig. \ref{Fig.Borel0d}), suggesting that the Stokes and anti-Stokes lines of the $d=1$ theory are same as the ones of $d=0$, i.e. $\theta=0$ and $\theta=\pi$ are Stokes lines and $\theta= \pm \frac{\pi}{2} $ are anti-stokes lines. 

For example, the contribution of the perturbative vacuum and the non-perturbative saddles with which it ``talks"  (i.e., the non-perturbative saddles which are in the same homotopy class as the perturbative vacuum) lead to a transseries of the form 
\bea 
 {\mc Z} (g^2 | m)=  \left\{ \begin{array}{lc} 
   \Phi_0 (g^2)  +     [{\cal I} \bar{\cal I}]_{-}    {\Phi}_{[{\cal I} \bar{\cal I}]} (g^2)    +       [{\cal I}^2  \bar{\cal I}^2]_{-}  
     {\Phi}_{[{\cal I}^2 \bar{\cal I}^2]} (g^2)    +   \ldots   \qquad& -\pi <  \theta< 0 \cr
    \Phi_0 (g^2)  +     [{\cal I} \bar{\cal I}]_{+}    {\Phi}_{[{\cal I} \bar{\cal I}]} (g^2)    +       [{\cal I}^2  \bar{\cal I}^2]_{+}  
     {\Phi}_{[{\cal I}^2 \bar{\cal I}^2]} (g^2)    +   \ldots      \qquad  & 0 < \theta < \pi  
\end{array} \right. 
\ea
where    $\Phi_0 (g^2)$ is the ordinary   Rayleigh-Schr\"odinger perturbative expansion, and  $ {\Phi}_{[{\cal I} \bar{\cal I}]} (g^2) $ is the perturbative expansion around the non-perturbative saddle, {\it etc}. Similar to the $d=0$ example, we find that the coefficient of the expansion around the
non-perturbative saddle at $\theta= 0^{\pm}$ is discontinuous:   
  $[{\cal I} \overline {\cal I}]_{\pm}  \sim \left( \log \frac{1}{g^2} - \gamma \pm  i \pi  \right) e^{-2 S_{\cal I}}$, since  
 $\arg(g^2) = 0$ is a Stokes line.   In fact, the purely imaginary ambiguous part  of the amplitude cures  the ambiguity associated by the 
non-Borel summability of the perturbation theory,  (exactly as in the $d=0$ example)   
 i.e.,  
 \bea 
 {\rm Im} \left(  {\cal S}_{0^\pm} \Phi_{0}+ [{\cal I} \overline {\cal I}]_{0^\pm}   {\rm Re} {\cal S}_0  {\Phi}_{[{\cal I} \bar{\cal I}]}   \right) =0   \quad \text{up to}   \;\;  \mc O(e^{-4S_I})   \;\;  
\label{a-c-3}
\ea  
 the  counter-part of the $d=0$ relation (\ref{a-c-2}). Such relations render  the path integral real and unambiguous on the Stokes line, also see \cite{A-S}.

 As the theory is analytically continued to 
 $\arg(g^2) = \pi^{-}$,  clearly all  $[{\cal I}^k \overline {\cal I}^k]$ terms become exponentially large, similar to what the $B$ saddle contribution does in the vicinity of    $\arg(g^2) = \pi^{-}$ in the $d=0$ example of Section \ref{sec:zero}.  Therefore,  
 we conclude that the path integration cycle of  the analytically continued  theory cannot be  the one associated with the dual theory. 
 
 However, as in the $d=0$ case, there exists a dual cycle, which is the uplift to $d=1$ of the  zero dimensional dual cycle.  Conversely, the $d=0$ limit can be obtained by dimensional reduction of the quantum mechanical model as $\beta\rightarrow0$. 

\begin{align}
 \label{0d-1d-cycles}
 \setcounter{MaxMatrixCols}{20}
\begin{matrix}
d=1:   \qquad   & \Gamma(\pi^{-}|m)  & \xleftarrow{\rm analytic  \; cont.}   	   & \Gamma(0^{+}|m)		 &	\xleftrightarrow{ \rm dual \; cycle} &         \tilde	\Gamma (\pi^{-}|m)		     \cr 
	      & \updownarrow     &     &  \updownarrow 	  						&    &\updownarrow  \cr
d=0:  \qquad    &  - {\cal J}_B(\pi^{- }) + {\cal J}_A (\pi^{-} ) & \xleftarrow{\rm analytic  \; cont.}    &  {\cal J}_A(0^{+}) -  {\cal J}_B (0^{+}) 	 &	\xleftrightarrow{\rm dual \; cycle} &          	{\cal J}_A(\pi^{- }) -  {\cal J}_C (\pi^{-})	     \cr 	          	          	
\end{matrix}
\end{align}
for which the partition function 
  \begin{eqnarray}
 \int_{\tilde\Gamma(\pi^{\mp}|m) } {\mathcal D} x \, e^{+ \frac{1}{|g|^2} \int_0^{\beta} d\tau \left( \frac{1}{4}  
\dot x^2
 +\, \sd^2( x |m) \right)
  } 
\label{path-int-3}
\end{eqnarray}
 does not have any exponential growth.   In fact,  $\int_{\tilde\Gamma(\pi^{\mp}|m) } {\mathcal D} x$ is an integration over the space of  purely imaginary paths due to the self duality
 \bea
 \tilde\Gamma(\pi|m)=i \,\Gamma(0|\mpr)
 \ea
 which is implemented by redefining $x(\tau)= i y (\tau)$ with $y(\tau)\in\mathbb R$ and $m\rightarrow\mpr$ in (\ref{path-int-3}). 
 The exponential of the action comes with  
 $e^{+ S[x]/|g|^2 }$ such that the integration is damped sufficiently fast, similar to  the familiar Gaussian integral, $\int_{- i \infty}^{+ i \infty} e^{ x^2} dx  < \infty$, over  
 $i \mathbb R$.

Note that the analytic continuation cycle obtained at $\theta=\pi$,  and the dual cycle,   are different. 
The  transseries  for a fixed value of $m$  for $g^2>0$, and $g^2 \rightarrow -g^2$  with the choice of the dual cycle, 
 are given by 
\begin{align}
 \label{resurgence-sd}
 \setcounter{MaxMatrixCols}{20}
\begin{matrix}
g^2>0:    &   \ldots \longleftrightarrow &    			-4S(m')   		 &	\longleftrightarrow&          	 -2S(m')  	       &\longleftrightarrow & 		 0 	    	   &  \longleftrightarrow   &	      2S(m) 	           &  \longleftrightarrow& 4S(m)&  \longleftrightarrow  \ldots \cr 
 		  &	      						  &			\updownarrow		 & 				          &       \updownarrow 	  	& 					   &\updownarrow&					        &\updownarrow  	    &   					 & \updownarrow&    	\cr 
 		  &	  \ldots \longleftrightarrow & { [{\cal G}^2 \bar{\cal G}^2]} & \longleftrightarrow &  { [{\cal G} \bar{\cal G}] }   & \longleftrightarrow & {\rm pert. vac.} &  \longleftrightarrow    & [{\cal I} \bar{\cal I}]    &      \longleftrightarrow&[{\cal I}^2 \bar{\cal I}^2] &      \longleftrightarrow \ldots   		\cr
		  &    						  &    	    \updownarrow		  & 					   &     \updownarrow     	        &					   &\updownarrow &					        &\updownarrow          &						   &\updownarrow &  	\cr  
 g^2<0:   &    \ldots \longleftrightarrow  &     		-4S(m) 	    		        & \longleftrightarrow &    -2S(m)  				 & \longleftrightarrow  & 0   			     &  \longleftrightarrow &   2S(m')  			   &   \longleftrightarrow&   4S(m') &      \longleftrightarrow \ldots        	\cr		
\end{matrix}
\end{align}
where $S(m)\equiv2 \sin^{-1}(m)/\sqrt{m\mpr}$. To recapitulate, 
 for $g^2>0$, the transseries parameters of the saddles with positive action $2nS(m)$ are turned on, while the ones of the negative action,  $-2nS(m')$, are turned off.     Taking $g^2 \rightarrow g^2e^{i \pi}$  by analytic continuation turns the ghosts (see \ref{eq:ghost}) into real instantons (see \ref{eq:inst}) with action 
 $S(m')$,  but they are not on the analytically continued cycle. Simultaneously, it also turns real instantons into ghost instantons  with action $-S(m)$, and  they are still on the analytically continued cycle leading to exponentially growing $e^{+{S(m)\over|g|^2}}$ terms. Therefore, the analytically continued theory  is an undesirable theory, with pathological exponential growth.  On the other hand, we know that the simple self-duality property of the potential maps the theory with a given elliptic parameter $m$ to itself with the change of $m\rightarrow\mpr$. 

 The resolution of this puzzle is that there also exists another  linear  combination of the  Lefschetz  thimbles, the dual cycle, on which instantons with action  $S(m')$ live. However, ghost-instantons with action $-S(m)$ do not  live on the dual cycle. The self duality is realized in the path integral by changing the cycle to the dual cycle. Since changing the original cycle to the the dual cycle at $\theta=\pi$ is a non-analytic process,  the original partition function and the dual partition  function are associated with different quantum phases (characterized by $m$ and $\mpr$) of a given Hamiltonian.  This is discussed in the next subsection.

\subsection{Quantum phase transition }
\label{QPT}
Consider the Hamiltonian 
\bea
 H_m = - \frac{d^2}{d \phi^2} + \frac{1}{g^2}\, \sd^2(g\,\phi|m)  \xrightarrow{g^2 \rightarrow -g^2}  H_{\mpr} = - \frac{d^2}{d \phi^2} + \frac{1}{g^2}\, \sd^2(g\,\phi|\mpr)  
  \ea
and its Hilbert space structure.  There are in fact infinitely many different theories associated with this Hamiltonian, and in order to 
give a concrete description of the Hilbert space, one needs to state  the identifications in the configuration space.  This is equivalent to gauging the global discrete symmetries, see below and 
Ref.~\cite{Unsal:2012zj} for more details. These definitions will be useful to give a simpler description of quantum phase transition.

{\it Infinite configuration space, $T_{\infty}$-model:}
If the configuration space is 
$g\,\phi \in \mathbb R$,  the potential leads to an infinite lattice stucture, and the low-lying modes form a Brillouin zone, with energy dispersion relation $E[(g^2|m), k]$, where   $k \in [- \pi, \pi]$ is momenta in Brillouin zone.  We call this the $T_{\infty}$-model, where $T_{\infty}$ is the discrete translation symmetry which commutes with the Hamiltonian. There are infinitely many states in the lowest-Brillouin zone, and their splitting is due to  instanton effects. 

 {\it Compact configuration space, $T_{N}$-model:} If we physically identify  $g\,\phi  \equiv g\,\phi +  N\times [2 \K(m)]$, the resulting system may be viewed as a particle on a circle with $N$-minima on the circle. We call this the $T_N$-model. The $T_N$-model is a gauged version of the $T_{\infty}$-model, in which sites with $N$-lattice spacing apart are physically identified. 
In the $T_N$-model, there are $N$-low lying, non-perturbatively split (by instanton effects) modes.  
The $T_1$ (particle on a circle, with single minima on the circle) has a unique ground state both in perturbation theory and non-perturbatively.  It is the $k=0$, the low-edge state of the $T_{\infty}$-model. 
For $T_2$ model, we have 
two-low lying modes, with  symmetric/anti-symmetric   wave functions.  These are the low and high-edge state of  the lowest band for the energy-dispersion relation in the 
the $T_{\infty}$-model. We call the splitting between these two modes as the 
{\it mass gap},  the energy required to excite the system from the ground state  $E_0$ to the first excited state $E_1$, and denote it as 
${\mathfrak M_{\mathfrak g}} (g^2|m) $. 
In the $T_{\infty}$ model, this is equal to the band-width for the energy dispersion relation  $E[(g^2|m),k]$, namely, 
\bea
\label{order}
{\mathfrak M_{\mathfrak g}} (g^2|m) &=&  E_{1} - E_{0}    \qquad {\rm in\,\, the}   \;\;T_2{-\rm model} 
\\
& = &  E[(g^2|m), k=\pi] -  E[(g^2|m), k=0]  \equiv  \Delta   E(g^2|m)  \qquad {\rm in\,\, the}  \;\; T_\infty{-\rm model}  \nonumber
\ea
Either expression in  (\ref{order})  is the order parameter to probe the quantum phase transition.  

{\it Particle on a circle with topological $\Theta$-angle, $T_{1}(\Theta)$-model:}
Finally, one can introduce a topological $\Theta$-angle in the $T_1$-model, and construct the  $T_1(\Theta)$-model. Still, this theory has a unique ground state  to all orders in perturbation theory, and also non-perturbatively. However, by feeding in momentum into the system, i.e., identifying the $\Theta$-angle with the Bloch momentum,  $\Theta \equiv k a$,  ($a$ is lattice spacing)
the ground state can be chosen to be any state of the lowest energy band. Functionally, the $\Theta$-dependence of the 
ground state energy is identical to the energy dispersion relation under this identification; i.e., 
\bea
\label{order-2}
 E[(g^2|m), ka]  \equiv  E[(g^2|m), \Theta]  
 \ea

Consider the process of taking $g^2 \rightarrow -g^2$  for a fixed value of $m$, by gradually decreasing $g^2$. The mass gap in the $T_2$-model, or equivalently the  band-width in the $T_\infty$-model,  is a one-instanton effect at leading order in weak coupling:
\bea
\Delta E(g^2|m) \underbrace{\longrightarrow}_{g^2 \rightarrow -g^2}  \Delta E(-g^2|m)= \Delta E(g^2|1-m) \; ,
\ea 
This implies a non-analytic behavior for the mass gap at $g^2=0$: 
\bea 
\label{gap-f}
\begin{array}{ll} 
- \frac{1}{\log {\mathfrak M_{\mathfrak g}}}  \sim \left\{ \begin{array}{ll}
 \frac{g^2}{S_{\cal I}(m)}    \cr 
  \frac{-g^2}{S_{\cal I}(1-m)}  
 \end{array} \right.    \qquad {\rm or}  \qquad
&
\frac{\partial}{\partial (g^2)} \frac{-1}{\log {\mathfrak M_{\mathfrak g}}}  \sim \left\{ \begin{array}{ll}
 \frac{1}{S_{\cal I}(m)},     &  \qquad   g^2 >0  \cr 
  \frac{-1}{S_{\cal I}(1-m)},     &  \qquad  g^2 <0  
 \end{array} \right.  
 \end{array} 
\ea
This non-analyticity of the mass gap is a quantum phase transition.
\begin{figure}[htb]
\includegraphics[scale=0.46]{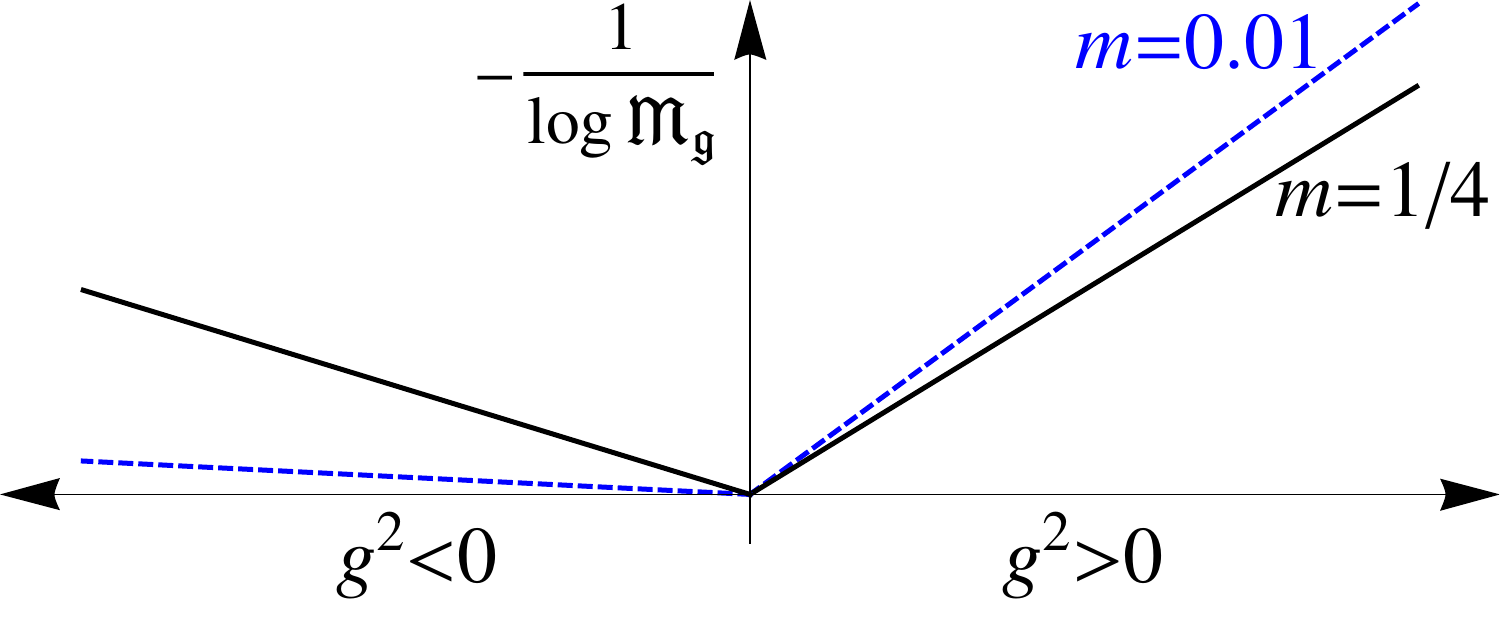}
\hspace{1.1cm}
\includegraphics[scale=0.46]{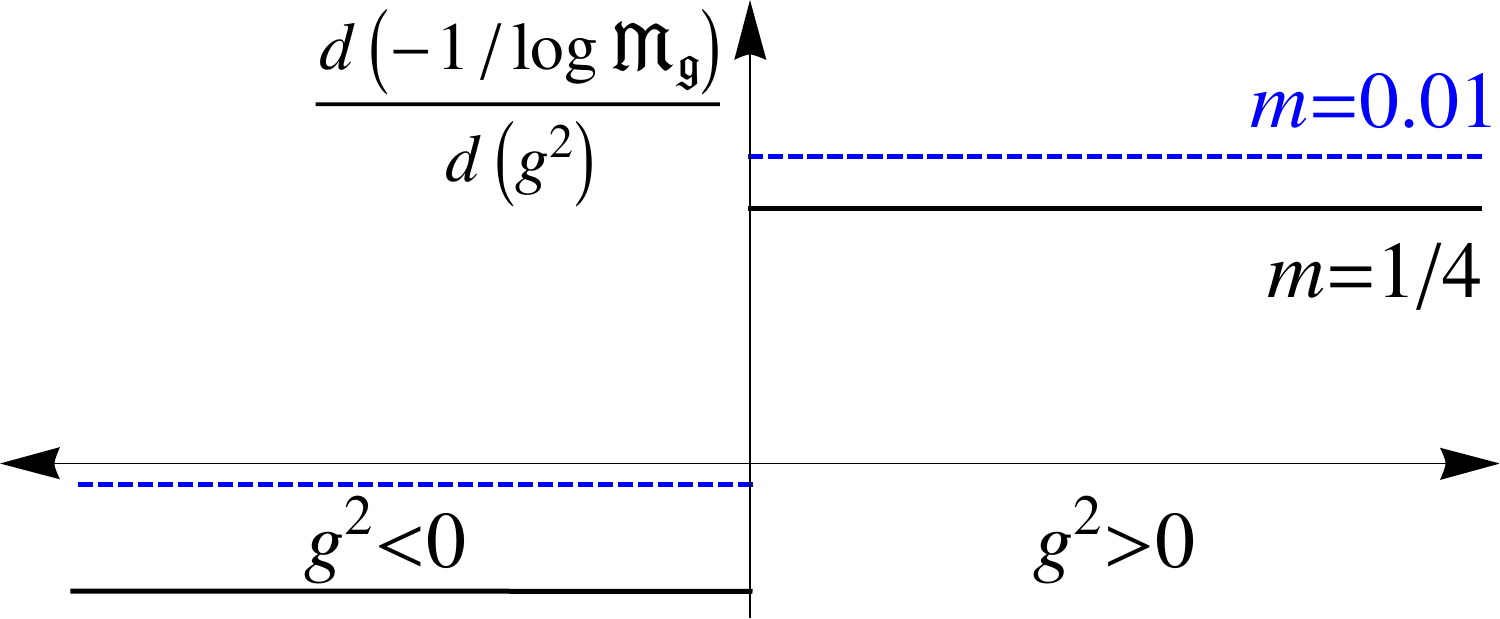}
\caption{The band-width/gap functions (\ref{gap-f}) as a function of $g^2$ for fixed $m$. The non-analytic behavior at $g^2=0$ is a quantum phase transition where the set of instantons contributing to the mass gap changes abruptly from the one with action $S(m)$ to the one with action $S(1-m)$. 
}
\label{Fig.QPT}
\end{figure}

 $\bm{ 0<m<1:}$ On the $g^2 >0$ side, the gap is due to the real instanton. Taking  
 $g^2 \rightarrow -g^2$, turns an instanton    with  positive action 
 into a ghost-instanton with negative action, and   the ghost-instanton  into a real instanton.  
  Therefore,  the action of the leading real instanton configuration contributing to the mass gap jumps abruptly from $S_{\cal I}(m)/g^2$ into $S_{\cal I}(m')/g^2$.  Therefore, the gap functions defined in (\ref{gap-f}) exhibits 
 a non-analyticity at $g^2=0$. 

 $\bm{ m=0:}$ On the $g^2 >0$ side, the gap is due to the real instanton, while on the $g^2 <0$ side, there are no real instantons. 
 In this case,  there is no band-formation (or  $\Theta$ angle dependence) for $g^2<0$, and these observables are  non-analytic at $g^2=0$.  
  
  Perhaps most importantly, the two quantum phases associated with the Hamiltonian for $g^2 > 0$ and $g^2 <0$ are not 
  associated with cycles which are related by analytic continuation,  the first two cycles  in  (\ref{0d-1d-cycles}). 
  Rather, the integration cycle on the  $g^2 <0$ side involves a non-analytic choice of integration cycle, the last cycle in (\ref{0d-1d-cycles}). 
   This means that different linear combinations of Lefschetz thimbles in path integrals  find an interpretation as  different quantum  phases associated with the  physical system. It would be useful to understand these relation in more  depth for general cases. 
\section{Comparison  of   $d=0$ and $d=1$ cases and Darboux's theorem}
\label{sec:Darboux}
In \S.\ref{sec:zero} and \S.\ref{sec:one}, we essentially mapped to the  complex Borel plane a  zero dimensional partition function and a
one dimensional quantum mechanics problem in the path integral formulation. The Borel transform, and the 
singularities of its analytic continuation to the Riemann surface, captured non-perturbative,  or equivalently,   large-order perturbative aspects.   In particular we gave a complete description of the ÒfirstÓ singularities on $ \mathbb R^+$ and   $ \mathbb R^-$.

For $d=1$ and $d=0$, the Borel plane structure and the saddles that contribute to the  perturbative expansion around the perturbative  vacuum are encoded into the following infinite and finite chains:
 \begin{align}\label{resurgence-0-1}
\begin{matrix}
\ldots \longleftrightarrow    { [{\cal G}^2 \bar{\cal G}^2]}     & \longleftrightarrow &  { [{\cal G} \bar{\cal G}] }    & \longleftrightarrow & {\rm pert. vac.} &  \longleftrightarrow & [{\cal I} \bar{\cal I}]   & 
      \longleftrightarrow  [{\cal I}^2 \bar{\cal I}^2]        \longleftrightarrow \ldots   \\
         &&\updownarrow && \updownarrow  &&\updownarrow & \\  
   & &{ C} & \longleftrightarrow &A&  \longleftrightarrow & B  &    
 \end{matrix}
\end{align}
The first chain is the zero topological charge  column of the graded resurgence triangle \cite{Dunne:2012ae}. 

It is  surprising to see that the resurgent asymptotic analysis of an ordinary integral (the zero mode of path integral) seems to be 
very similar to the analysis of  path integration,  which is   infinitely many ordinary integrations.  The simplicity of this generalization is striking and unexpected.  
The form of asymptotic growth of the expansion coefficients in perturbation theory in $d=0$  (\ref{eq:d0-leading}) and $d=1$  (\ref{eq:answer})  exhibit the same pattern  modulo the replacements 
\bea
S_B \longleftrightarrow  S_{\cal I \bar{\cal I}}, \qquad \qquad S_C \longleftrightarrow  S_{\cal  G \bar{\cal G}}
\label{singularity}
\ea 
as also indicated in (\ref{resurgence-0-1}).

It is  important to emphasize that the mapping is {\it not}  $S_B \longleftrightarrow  S_{\cal I}, ;\; S_C \longleftrightarrow  S_{\cal  G }$, despite the fact that ${\cal I} $ or $ {\cal  G }$ are the leading saddles of the path integration, and their action is half that of the   $[{\cal I} \bar{\cal I}]$, $[{\cal G} \bar{\cal G}] $ saddles.  This aspect differs between the $d=0$ and $d \geq 1$ cases, where in $d=0$, $B$ and $C$ are the leading saddles, while in  $d=1$,   $[{\cal I} \bar{\cal I}]$,  and $[{\cal G} \bar{\cal G}] $ are not. However, 
 $[{\cal I} \bar{\cal I}]$  and $[{\cal G} \bar{\cal G}] $  are  the  leading saddles that can mix with perturbation theory around the perturbative vacuum. The way to see this is explained in Ref.\cite{Dunne:2012ae}.  Introduce a topological $\Theta$ angle, which is invisible in perturbation theory, but the  instanton amplitude depends on it. Consequently,  perturbation theory can {\it never}  cancel its ambiguities against the  instanton sector,    but it can happily  do so against $[{\cal I} \bar{\cal I}]$ events (which are, of course,  $\Theta$-independent). The same logic also holds for  ghost-instantons  vs. their pairs, as can be seen by using analytic continuation.   The instanton sector  will form an infinite resurgence chain of its own, 
in  parallel with (\ref{resurgence-0-1}), given by 
    \begin{align}\label{resurgence-inst.}
\begin{matrix}
\ldots \longleftrightarrow    { [{\cal I}{\cal G}^2 \bar{\cal G}^2]}     & \longleftrightarrow &  { [ {\cal I} {\cal G} \bar{\cal G}] }    & \longleftrightarrow &  [{\cal I} ] &  \longleftrightarrow & [{\cal I}^2 \bar{\cal I}]   & 
      \longleftrightarrow  [{\cal I}^3 \bar{\cal I}^2]        \longleftrightarrow \ldots   \\
    \end{matrix}
\end{align}
The Borel plane singularity structure that one would extract from (\ref{resurgence-0-1})  and  (\ref{resurgence-inst.}) are essentially  the same, meaning that perturbation theory around the perturbative vacuum  and around the instanton background are also  intimately related.

We now address the underlying reason for the similarity between the $d=0$ expression (\ref{eq:d0-leading}) and the $d=1$ expression (\ref{eq:answer}).    Why is this so?

A theorem by Darboux  seems to be at the heart of this similarity. 
Darboux's  theorem states that the late terms in the Taylor expansion of an ordinary function   are determined by the behavior of the function in the vicinity of its singularities, poles or branch points. See \cite{dingle} for a lucid explanation.  As a simple example, let $BP_0(t)$ denote the Borel transform of the perturbation expansion around the perturbative vacuum, a function with singularities at $\{b_1, b_2, \ldots\}$,  where $b_m$ is an increasing  set on $\mathbb R^{+}$,   and  at $\{c_1, c_2, \ldots\}$,  where $c_m$ is a decreasing set on $\mathbb R^{-}$. Then, Darboux's theorem states that the late terms in the Taylor expansion around $t=0$ are universal and dictated by $b_1, c_1, \ldots $ {\it etc.}.  Inserting this result into the directional Borel sum,   
${\mathbb B}_{n, \theta}(\lambda) = \int_0^{\infty \cdot e^{i\theta}} BP_n(t\lambda) e^{-t} dt$, 
we observe  that the asymptotic factorial growth of the large orders of perturbation theory is determined by the singularity locations in the  Borel plane. 
 The main  difference between the $d=1$ path integral  and the $d=0$ ordinary integration  is that the former has infinitely many singularities, both on $\mathbb R^{+}$ and  $\mathbb R^{-}$, whereas the latter has only one-singularity in each direction. This comparsion is  shown in  Fig.~\ref{Fig.Borel}.    This is a typical situation, since the Schwinger-Dyson equation for the $d=0$  case is a finite order ODE (third order in our case), whereas the the Schwinger-Dyson equations for the $d=1$  case is  infinite order. 
 One can interpret the contribution of the infinitely many saddles as a manifestation of a {\it functional} generalization of  Darboux's theorem. 
 
 However, at least at sufficiently weak coupling,  the effects of the singularities farther away from the origin, such as    $[{\cal I}^2 \bar{\cal I}^2]$  and $[{\cal G}^2 \bar{\cal G}^2] $ singularities,  are exponentially suppressed with respect to the effect of the leading ones,    $[{\cal I} \bar{\cal I}]$,  and $[{\cal G} \bar{\cal G}] $, respectively. Roughly, (omitting order one numbers irrelevant to the main point), we can
 express 
 \begin{align}
 & a_n \sim a_n^{(1)} \left[ 1+ O\left(\frac{1}{n}\right) \right] + a_n^{(2)} \left[ 1+ O\left(\frac{1}{n} \right)\right]    +  \ldots   
 \end{align}
 where the terms in the square brackets come about from the perturbative fluctuations around the $[{\cal I}^k \bar{\cal I}^k]$  and $[{\cal G}^k \bar{\cal G}^k]$  saddles. The growth in the late terms associated with these saddles are
 \bea
&  [{\cal I} \bar{\cal I}]  \longleftrightarrow a_n^{(1)}  \sim \frac{n!}{(2 S_I)^n},  \qquad \qquad \; \; [{\cal I}^k \bar{\cal I}^k]  \longleftrightarrow a_n^{(k)}  \sim \frac{n!}{(2k S_I)^n}\; , \cr
&[{\cal G} \bar{\cal G}]  \longleftrightarrow a_n^{(1)}  \sim \frac{(-1)^n n!}{(|2 S_{\cal G}|)^n}, \qquad \qquad [{\cal G}^k \bar{\cal G}^k]  \longleftrightarrow a_n^{(k)}  \sim \frac{(-1)^nn!}{(|2k S_{\cal G}|)^n}\; 
\label{suppress}
 \ea 
Thus, the effect of the higher-correlated saddles  [coming with a factor $k$ in (\ref{suppress})] are suppressed by an exponential  factor of  $e^{-n \log(2k)} $ with respect to the leading ones,   in the late terms of  perturbation theory, also see \cite{Argyres:2012vv}. 
  Thus, modulo the replacement (\ref{singularity}), 
  the mathematics of the Borel transform of the path integral taking into account the  leading saddles on $\mathbb R^{+}$ and  $\mathbb R^{-}$      and the one of ordinary integral are essentially the same.   
 
 \section{Conclusion and prospects}

In this work, we studied the implications of resurgence theory for quantum mechanical systems in the path integral formulation.  We considered a physical system  which possesses both instantons and ghost-instantons (real and imaginary saddles). We showed that regardless of being included in the path integration cycle or not, {\it all} the saddles contribute to the large order behavior of the perturbative expansion, and the dominant contribution comes from those that are associated with the singularities on Borel plane closest to origin. In the particular model we studied, the associated singularities in the Borel plane are located in two directions: $\theta_1=0$ and $\theta_2=\pi$. 

\begin{figure}[htb]
\begin{center}
\includegraphics[scale=0.6]{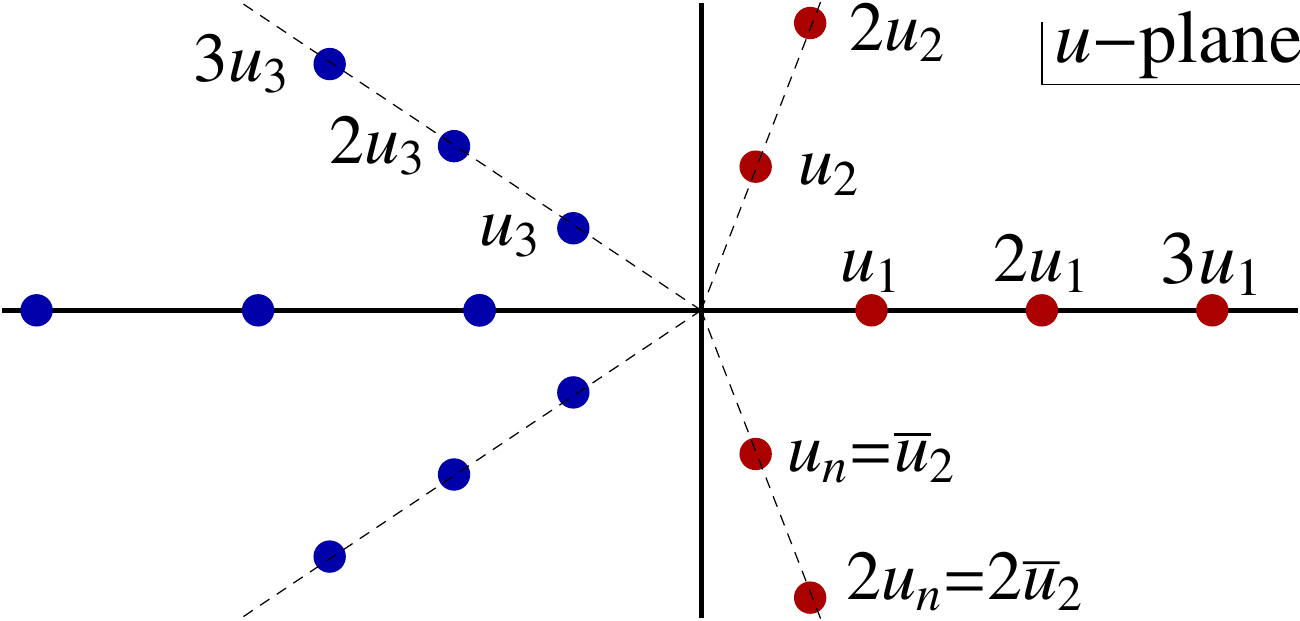}
\end{center}
\caption{Borel plane structure for a generic quantum mechanical system, quantum field theory in the general case. There are singularities along possibly a finite set of rays, emanating from the origin. 
}
\label{Fig.borel-gen}
\end{figure}

 For more general  potentials, we may also have singularities lying along an arbitrary number of  rays  $\theta_i,  \; i=1, \ldots n$, where $\theta_i$ is the argument of the  generalized instanton action.  We expect in general that in quantum mechanics, quantum field theory, as well as string theory and gravity, the typical picture of the Borel plane structure to look like Fig.~\ref{Fig.borel-gen}.  
 In general, the entire generalized instanton sector will govern the large-order asymptotics of perturbation theory around the perturbative vacuum, as well as any non-perturbative sector.  Contrary to what was believed to be a problem in the past,  having a non-Borel summable series with isolated singularities on  $\mathbb R^{+}$ is not a problem. 
  If the Borel transform of the perturbative expansion satisfies the ``endless analytic continuation" property (i.e., if the set of all singularities in all Riemann-sheets do not form any natural boundaries), then such systems can be studied using the resurgence formalism.   
 The resurgence theory of  \'Ecalle gives a  natural framework for describing the connections between the large-orders in perturbation theory and 
known and unknown (to be explored) saddles for a wide range of systems.  

Furthermore, we  showed that in addition to analytical continuation, there is an alternative way to define the theory for negative couplings. This alternative definition is achieved by defining the path integral for negative couplings on a different integration cycle (Lefschetz thimble)  than for positive couplings. In fact, in our particular model they belong to different homotopy classes, and therefore they cannot be connected via analytical continuation or Stokes phenomenon. We showed that this non-analyticity is related with a quantum phase transition. Also, as opposed to analytical continuation, the alternative definition replaces  real instantons with ghost instantons, and for negative couplings these objects do not have exponentially large contributions to the partition function, and lead to a well defined theory.

 The analytic continuation of path integration by itself has received some interest recently; see for example \cite{Jaffe:2013yia, Witten:2010cx, Harlow:2011ny, Guralnik:2007rx, Ferrante:2013hg}.  The connection between analytic continuation  of path integral and resurgence theory is much less explored  \cite{Garoufalidis}. 
 Our results suggest that the delf-duality respecting integration cycles  in the current example  can only be reached non-analytically, 
 and results in different quantum phases. 
 
 % non-analytic continuations of QFTs to negative/complex couplings which might lead to novel quantum phases and non-perturbative objects akin to the ghost insantons we discussed in this paper.

In  light of these observations, we conclude by remarking on a few possible directions and relations to some other works. 

\vspace{5mm}

{\bf  Complex saddles and QFT:}
One may wonder what our  analysis in quantum mechanics implies for general QFTs?  
It has recently been understood that IR-renormalons  \cite{'tHooft:1977am, Beneke:1998ui}  (singularities along $\mathbb R^{+}$) 
 in asymptotically free quantum field theories can be continuously connected to semi-classical configurations  involving neutral bions 
(these are pairs of a fractional BPST  instanton anti-instantons), in a semi-classical limit of  circle-compactified QCD and  ${\mathbb C}{\mathbb P}^{N-1}$  theories \cite{Argyres:2012vv, Dunne:2012ae}.  
It is also widely known that at least in some infrared-free theories, the IR-renormalons  (singularities along $\mathbb R^{-}$ now) are associated with complex instanton saddles \cite{'tHooft:1977am}. 
It would be interesting to understand the nature of the  general singularities in the Borel plane. 
 Complex saddles also occur after Wick rotation of path integrals (see for example \cite{Tekin:1998qy,Lavrelashvili:1989he,Alexanian:2008kd}). This is a somewhat different phenomenon from that studied in this paper, but it is likely that resurgent analysis will also be relevant to these problems.

\vspace{5mm}

{\bf Systematizing the  plethora of non-self-dual real saddles and complex saddles:} 
Historically,  real self-dual instantons (satisfying BPS bounds) have received the most attention in the QFT literature. However,  the world of non-BPS, 
yet semi-classically calculable/understandable saddles is proving to be equally, and sometimes even more,  interesting. 
There are many theories in which the homotopy group is trivial, yet, non-perturbative saddles can be classified according to resurgence theory and weak coupling methods \cite{Cherman:2013yfa}.  In theories with nontrivial homotopy group, resurgence theory and semi-classical calculability give a  refined classification of the non-perturbative saddles, which goes beyond topological arguments. 

\begin{itemize} 
\item {Recent work has shown that non-selfdual saddles play as prominent, sometimes even more important role, then the instantons, see  \cite{Argyres:2012vv, Dunne:2012ae,Dabrowski:2013kba}, and references therein, e.g.,  the role of the magnetic/charged  bions in mass gap, and 
the role of neutral bions in center-stability and in connection to renormalons).  These non-selfdual saddles are under analytic control in weak coupling. } Here, we  emphasized the role of complex saddles and complex  non-self-dual saddles (which can be realized in different Lefschetz thimbles). 
 \item{  Complex saddles may also be of importance in quantum gravity in dS space \cite{Banerjee:2013mca}. Similarly to our observation, Ref. \cite{Banerjee:2013mca} shows that whether or not there are exponentially growing terms in the partition function depends on the path integration cycle. }
 
\item{ In matrix model/string perturbation  theories,  it seems  plausible  that the terms in the trans-series exhibiting exponential growth  \cite{Aniceto:2011nu}  may be associated with some D-brane analogs of the ghost-instantons in QM\footnote{Here we do not refer to the ghost D-branes introduced in \cite{Okuda:2006fb}, but a more general set of objects whose actions are not necessarily equal to the negative of the D-brane action. }; see also \cite{Klemm:2010tm}.   At negative couplings, such objects may become real and vice versa.  }
\item
It would be worthwhile to understand more deeply the relation of resurgence theory to interesting recent work in lattice gauge theory  \cite{Cristoforetti:2012su,Bauer:2011ws}, and also to recent numerical results for resummation and non-perturbative effects in QM systems \cite{wosiek}.

\end{itemize}

\vspace{5mm}

 {\bf Connecting weak to strong coupling:} The transseries is the asymptotic expansion of a unique resurgent function at weak coupling.  In quantum mechanics of periodic potential, it is believed that there is no phase transition as the coupling is dialed from small to large.  
 It would be very interesting to obtain the strong coupling expansion  from the weak coupling expansion. 
  We have some numerical evidence in this direction, that the weak coupling transseries expansion with the inclusion of the asymptotically small terms and strong coupling expansion essentially merges at {\it intermediate, finite} coupling,   but so far, we could not give a  direct analytic proof \cite{du:qm}.  
 
 Another interesting theory in the weak-strong coupling  context is ${\cal N}=4$ SYM.  Ref.\cite{Argyres:2012vv} gave evidence for the absence of singularities on positive real axis, ${\mathbb  R}^+$ in Borel plane, i.e, for Borel summability along   ${\mathbb  R}^+$. 
  Namely, the ambiguity associated with  topological  molecules such as magnetic bion-anti-bion correlated pair  which lead to semi-classical realization of IR-renormalon in QCD(adj) on  $\mathbb R^3 \times  S^1_L$
does not exist in  ${\cal N}=4$ SYM, hence by continuity on   ${\mathbb R^4}$.  On ${\mathbb R^4}$, there is also a  strong belief that  ${\cal N}=4$ SYM has  no phase transition as a function of the 't Hooft coupling, and  there also exist  evidence of a strong-weak or S-duality.  Both of these restrictions may be useful to construct S-duality invariant resurgent transseries for  ${\cal N}=4$ as also emphasized in \cite{A-S}. (This would be the counter-part of   self-duality invariant resurgent transseries, a property  that helped  our analysis.)  For related recent work, see \cite{Beem:2013hha}.

\vspace{5mm}

\vskip0.2cm

\acknowledgments
We are indebted to  R. Schiappa for inspiration and communications. We thank   P. Argyres, E. Poppitz, M. Anber, D. Ferrante, J. Greensite and M. Marino    
 for useful discussions and communications. 
We acknowledge support from the DOE grants  DE-FG-88ER40388 (GB),  DE-FG02-92ER40716 (GD), and  DE-FG02-12ER41806 (M\"U).

%%%%%%%%%%%%%%%%%%%%%%%%%%%%%%%%%%%%%%%%%%%%%%%%%%%%%%%%%%%%%%%%%%%

\end{document}